\documentclass[aps,prl,reprint,superscriptaddress,noshowpacs,titlepage,10pt]{revtex4-1}
\usepackage{graphicx}
\usepackage{amsmath}
\usepackage{bbm}

\begin{document}

\title{Population stability risks and biophysical benefits of cell-cell fusion in macrophage, osteoclast, and giant multinucleated cells}

\author{Jesse L. Silverberg}
\email[]{jesse.silverberg@wyss.harvard.edu, jesse.silverberg@gmail.com}
\affiliation{Wyss Institute for Biologically Inspired Engineering, Harvard University, Boston MA 02115}
\affiliation{Department of Systems Biology, Harvard University, Boston MA 02115}

\author{Pei Ying Ng}
\email[]{peiying\_ng@hsdm.harvard.edu}
\collaboration{Co-first authors}
\affiliation{Harvard School of Dental Medicine, Harvard University, Boston MA 02115}

\author{Roland Baron}
\email[]{Roland\_Baron@hsdm.harvard.edu}
\affiliation{Harvard School of Dental Medicine, Harvard University, Boston MA 02115}
\affiliation{Department of Medicine, Harvard Medical School, Boston MA 02115}

\author{Peng Yin}
\email[]{Peng\_Yin@hms.harvard.edu}
\collaboration{Co-last authors}
\affiliation{Wyss Institute for Biologically Inspired Engineering, Harvard University, Boston MA 02115}
\affiliation{Department of Systems Biology, Harvard University, Boston MA 02115}

\begin{abstract}
Plant and animal cells are commonly understood as acquiring specialized functions through differentiation and asymmetric division.  However, unique capabilities are also acquired when two or more cells fuse together, mixing cytoplasmic and genetic material.  Here, we combine imaging experiments with biophysical modeling to perform the first risk-benefit analysis of cell-cell fusion.  On one hand, we find fusion introduces an intrinsic instability to the population dynamics.  On the other hand, we measure an unusual physiological scaling suggesting these cells grow substantially larger at lower energetic costs.  Further analysis of the cytoskeleton finds a size-associated phase separation of F-actin that self-organizes multinucleated cell phenotypes.  
\end{abstract}

\maketitle

A contemporary understanding of cell-cell fusion is largely based on signaling pathways and protein structure\cite{ogle2005biological,podbilewicz2014virus,chen2005unveiling,chen2007cell,zito2016united}.  This biomolecular perspective has emerged from decades of studies working to unravel various aspects of how two cells can fuse membranes and cytoplasm to become one.  Developmental biology offers a number of examples for understanding fusion including the formation of zygotes, placenta, \textit{C. elegans} epidermis, muscle fibers, hybridomas\cite{chen2005unveiling}, osteoclasts\cite{oren2007cell,bar2007osteoclast,vignery2005macrophage,ishii2008osteoclast,vignery2000osteoclasts}, and multinucleated giant cells.  In each of these cases, fusion is essential for the development of specialized functions.  Other examples where fusion plays a narrower but still crucial role include stem cell reprogramming\cite{ogle2005biological,chen2005unveiling}, chromosomal instability, and tumor cell metastasis\cite{ogle2005biological,duelli2007cell,bastida2016dark}.  In these cases, we find cell-cell fusion contributing to the emergence of disease through undesirable interactions and horizontal transfer of cytoplasmic or genetic material.  While this brief sampling of developmental and biomedical examples does not cover the full scope of investigations into how cells fuse, fundamental biophysical questions emerge.  For example, what are the potential benefits of cell-cell fusion as a path to specialized function?  What are the potential risks? 

\begin{figure}
\includegraphics[scale=.9]{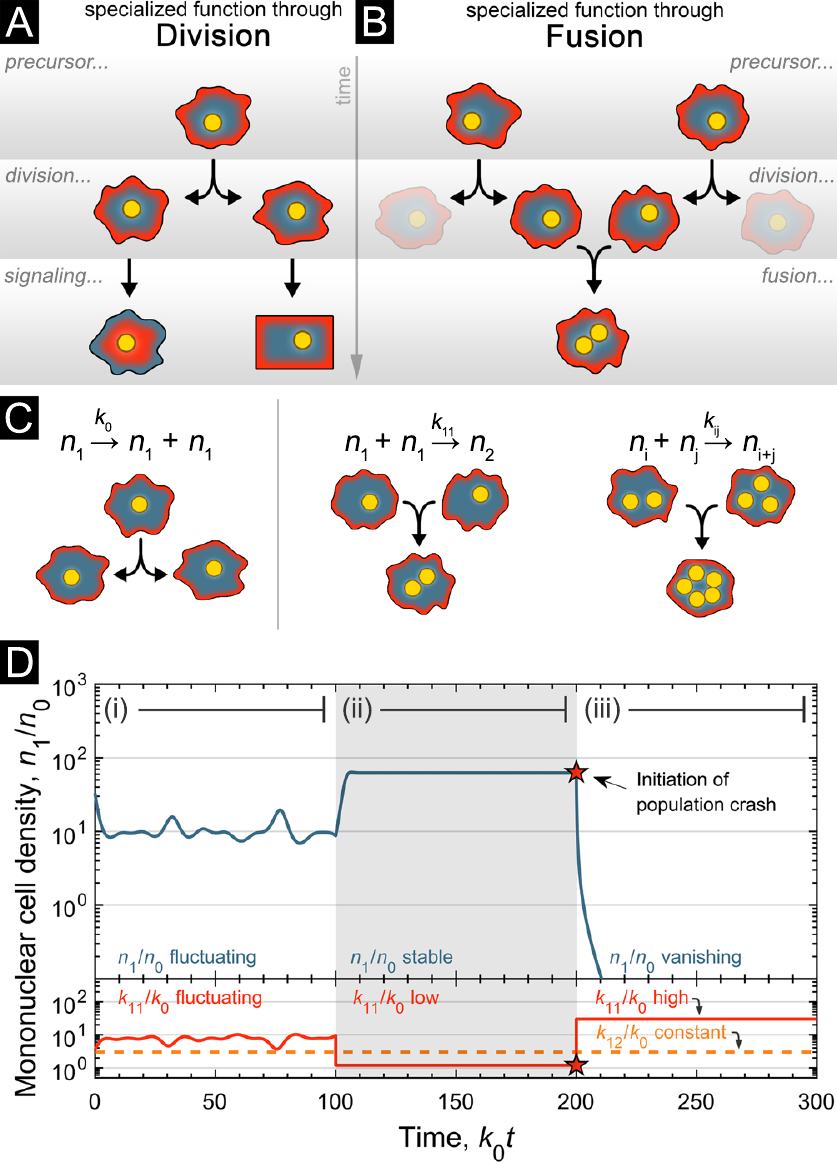}
\caption{Cell-cell fusion leads to irreversible population dynamics. (A) Many cell types are generated by asymmetric division.  (B) Cell-cell fusion leads to multinucleate cell types with unique biological functions.  (C) Fusion can be modeled with mass action rate equations: (left) division of precursor cells, (middle) fusion of two mononuclear cells, and more generally (right) fusion of two cells with $i$ and $j$ nuclei into a cell with $i+j$ nuclei. (D) Simplified ODE fusion model shows population dynamics (blue, solid) of mononuclear cell density $n_1$ (i)~fluctuates, (ii) remains stable, or (iii) crashes according to the fusion rate $k_{11}$ (red, solid).  Normalizations: cell density by $n_0$ from experiments; time $t$ by mononuclear division rate $k_0$. Fusion rate $k_{12}$ (orange, dashed) held constant.} 
\label{fig1}
\end{figure}

{\it Fusion Risks for Population Stability.} In fact, fusion generally introduces the potential for unstable cell population dynamics.  Unlike asymmetric division (Fig.~\ref{fig1}A), which \textit{adds} to the total cell population, fusion creates multinucleated cells at the \textit{expense} of a precursor cell population (Fig.~\ref{fig1}B).  Thus, if precursor cells divide faster than they fuse, the multinucleated cell population will steadily rise.  Conversely, if fusion happens more frequently than division, the precursor population will crash.  Putting specific details of biochemistry, apoptosis, genetic rescue\cite{jacome2019developmental}, and regulatory signaling aside, we see fusion appearing in any context is accompanied by an intrinsic and irreversible balance between population growth and loss.  This observation is potentially relevant for Paget's disease\cite{menaa2000enhanced} and giant-cell tumors\cite{nascimento1979primary}, where irregularities in the population levels of fusion-derived cell lines are well-known, but the full mechanisms driving these dynamics are yet to be understood.

Fusion's effect on population dynamics can be quantitatively investigated using a second-order rate-equation model studied with a combination of numerical and analytical techniques\cite{dai2012generic, dai2015relation}.  Representing the density of cells with $i$ nuclei as $n_i$ (Fig.~\ref{fig1}C), then precursor cell division happens at a rate $k_0$ according to $n_1 \rightarrow n_1 + n_1$.  Similarly, fusion between two cells with $i$ and $j$ nuclei happens in parallel at a rate $k_{ij}$ according to $n_i + n_j \rightarrow n_{i+j}$.  Combinations of $i$ and $j$ produce an infinite series of differential equations containing linear terms of the form $k_{ij} n_i n_j$ with $i \ne j$, quadratic terms of the form $k_{ii} n_i n_i$, and a parameter space of rate constants that grows quadratically with the index used to denote the number of nuclei.  Taking a numerical approach to solving these expressions by stochastically sampling parameter values reveals the steady-state behavior is either: (i) a linearly growing total cell population, or (ii) a complete loss of the mononuclear precursors (SM, Fig.~S1A).  In fact, nearly 80\% of the parameter space produces outcomes where the mononuclear precursor cell population was irreversibly wiped out by fusion.  To further understand the risks introduced by fusion, we examine a simplified analytic version of the rate-equation model with a maximum of 3 nuclei/cell (SM, Figs.~S1B and S2).  In this reduced form, there are two parameters to consider: the relative fusion rate $k_{12} / k_0$, which we hold constant, and the relative fusion rate $k_{11} / k_0$, which we vary.  When we allow $k_{11}/k_0$ to fluctuate randomly about an average value, the precursor cell population $n_1$ exhibits a similar fluctuation about a corresponding average density (Fig.~\ref{fig1}D, region i).  When we reduce $k_{11}/k_0$ and hold it constant, the precursor population $n_1$ plateaus at a higher equilibrium value and remains stable due to the new relative balance of fusion and division (Fig.~\ref{fig1}D, region ii).  When we increase $k_{11}/k_0$ to a higher value, then the density of precursor cells rapidly vanishes in a population crash (Fig.~\ref{fig1}D, region iii).  Remarkably, only a 4-fold variation in the fusion rate (Fig.~\ref{fig1}D, compare regions i and iii) was required for this irreversible outcome.  

{\it Big Data Imaging Experiments.} Appreciating the risks to population stability implied by the mathematics of cell-cell fusion, we used an experimental \textit{in vitro} system of mouse macrophage cells (SM) to investigate fusion's potential benefits.  When induced with RANKL stimulation\cite{rahman2015proliferation} (Fig.~S3), these macrophage precursor cells differentiate and fuse into osteoclasts with 2 to 20 nuclei, and eventually form giant cells with 100 or more nuclei.  Functionally, osteoclasts degrade mineralized bone during skeletal maintenance, repair, and remodeling.  Giant cells are less well understood\cite{miron2016giant}, but play a role in foreign body rejection and accumulate at infection sites.  Remarkably, these multinucleated cells grow $10$ to $10,000 \times$ larger than their mononuclear precursors (SM), leading to significant technical challenges in data acquisition, processing, and visualization.  Considering precursor macrophages are $\sim 10 \ \mu$m, this scale suggests a minimum $40\times$ objective for microscopy imaging.  With the dramatic increase in areal size we can anticipate large multinucleated cells approaching $\sim 0.5$ mm across.  However, even larger fields of view must be captured to obtain unbiased statistics.  Imaging with 150 nm pixels covering a $5 \times 5$ mm$^2$ region of interest therefore implies gigapixel-scale microscopy images.  For a single 3-color confocal experiment with 35 $z$-planes, which compensates for slightly tilted samples over large scales where auto-focusing techniques can fail, we anticipate raw image data generated at a rate of $(16$ bits/px) $\times \ (1.1 \times 10^9$ px/image) $\times \ (35$ images/color) $\times \ (3$ colors/sample) $= 231$ GB/sample.  As such, the wide range of physical length scales generated by fusion rapidly lead to challenges in Big Data imaging\cite{meijering2016imagining,knothe2016organ,husz2012web,pietzsch2015bigdataviewer,maree2016collaborative,bria2015open}.

\begin{figure*}
\includegraphics[scale=.9]{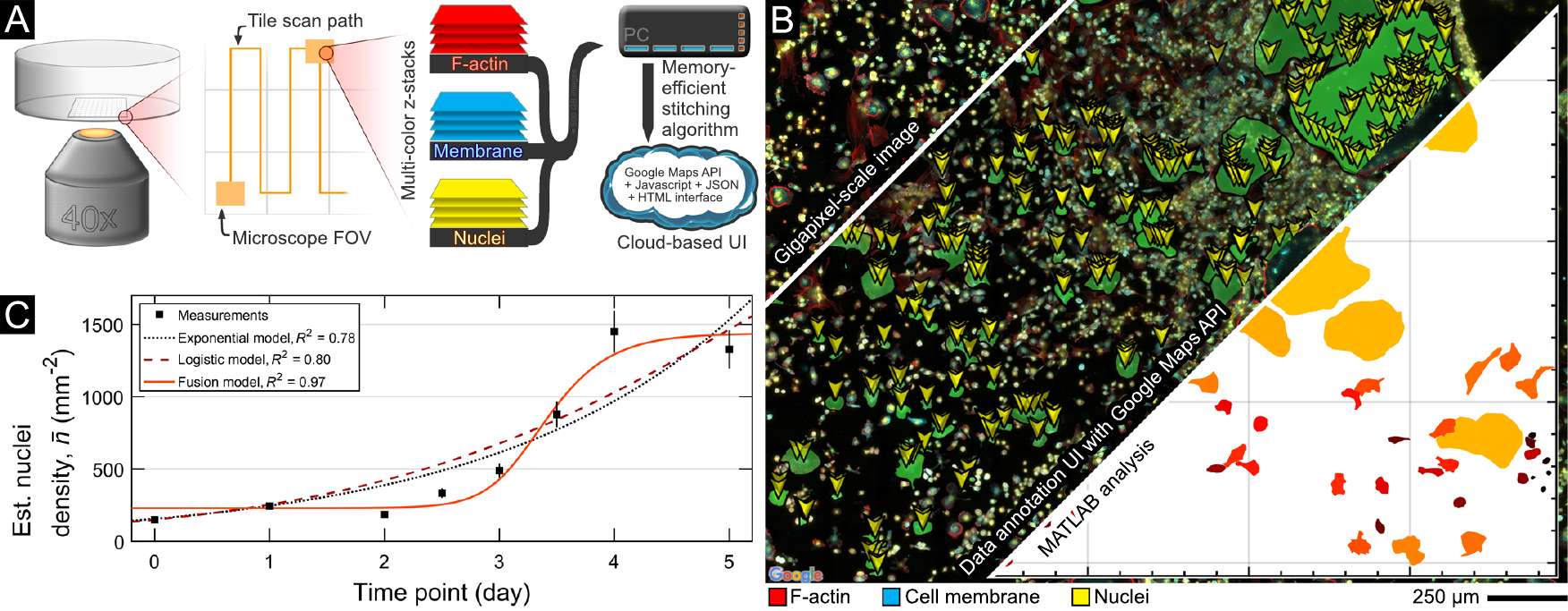}
\caption{Experimental pipeline for gigapixel-scale image acquisition and processing. (A) Schematic of data acquisition, Big Data processing, and cloud-based viewing.  (B) A small region of a gigapixel-scale image (top-left), annotations generated with the Google Maps web-based data viewer (middle), and extracted annotations plotted for analysis (bottom-right).  This field of view includes numerous multinucleated cells of various sizes.  Annotations color code: each green shaded region is a multinucleated cell; yellow arrows denote individual nuclei. (C) Estimated nuclei density measured from image data as a function of fixation time point (dots).  Conventional population dynamics models including exponential (dotted) and logistic (dashed) fail to reasonably capture the data.  Best-fit of analytic solution to fusion dynamics population model captures the data trends with high accuracy (solid).}
\label{fig2}
\end{figure*}

To experimentally study cell-cell fusion, we plated macrophage cells with an initial density of 150 cells/mm$^2$, induced fusion with RANKL stimulation, and fixed samples after a controlled period of time for imaging (SM).  Each sample was then tile-scanned with 3D spinning disk confocal microscopy to image the F-actin, membrane, and nuclei of all cells within a mm-scale field of view (SM).  These markers allowed us to distinguish clusters of mononuclear cells from multinucleated cells, since the membrane of cells within clusters has a distinct fluorescent signature observable in both 3D reconstructions and 2D projections.  We then developed a software pipeline using custom code and memory-efficient algorithms to process the raw microscopy data (SM; Fig.~\ref{fig2}A).  This pipeline synthesized individual tile-scanned images into the anticipated gigapixel-scale imagery, which was slow to render on-screen and impractically large for analysis.  We then used the same software tool to reprocess the data into a format compatible with the Google Maps JavaScript API, effectively replacing the default atlas imagery with microscopy data.  Subsequently, we developed a web-based viewer that (i) provided near-instantaneous rendering at any desired zoom, (ii) enabled dynamic switching between all color channel combinations, and (iii) offered a variety of measurement tools accessible from any internet-connected device with a web-browser (SM; Fig.~\ref{fig2}B).  Data annotations and measurements were stored in the cloud and could be saved, shared, or downloaded for further analysis.  We then used this Big Data imaging technology platform to quantify previously unknown population-scale trends driven by fusion.

Our data visualization and analysis tool enabled us to identify individual cell nuclei, note their locations, and compute the number of nuclei within the field of view.  This measurement allowed for an estimate of the nuclei density $\bar{n}$ at time points $t$ corresponding to 0, 1, 2, 2.5, 3, 3.5, 4, and 5 days post RANKL stimulation (Fig.~S4).  The measurements of $\bar{n}$ were relatively steady for the first 2 days, and then exhibited a rapid uptick to a new steady-state value that was reached on or near day 4 (Fig.~\ref{fig2}C, black squares).  Using MATLAB to compute a best-fit to the data, we found conventional exponential and logistic models of growth (SM) perform similarly in terms of their quality of fit ($R^2 = 0.78$ and $0.80$, respectively).  However, the exponential model predicts unfettered growth, which is inconsistent with our observations (Fig.~\ref{fig2}C, dotted black line), and the logistic model fails to capture the delayed-onset of the population plateau as indicated by the horizontally and vertically shifted sigmoidal shape of the $\bar{n}(t)$ measurements (Fig.~\ref{fig2}B, dashed red line).  As such, both models similarly under- and over-shoot the dynamics despite being best-fits to the data.  In contrast, the simplified rate-equation model of cell-cell fusion captures both the delayed uptick and plateau in population density when fit to the measurements ($R^2 = 0.97$; Fig.~\ref{fig2}B, solid line; SM).  Thus, the relative balances between fusion and division that revealed a theoretical possibility of population crashes also has a capacity to better account for empirical trends in the population data.

\begin{figure*}
\includegraphics[scale=.9]{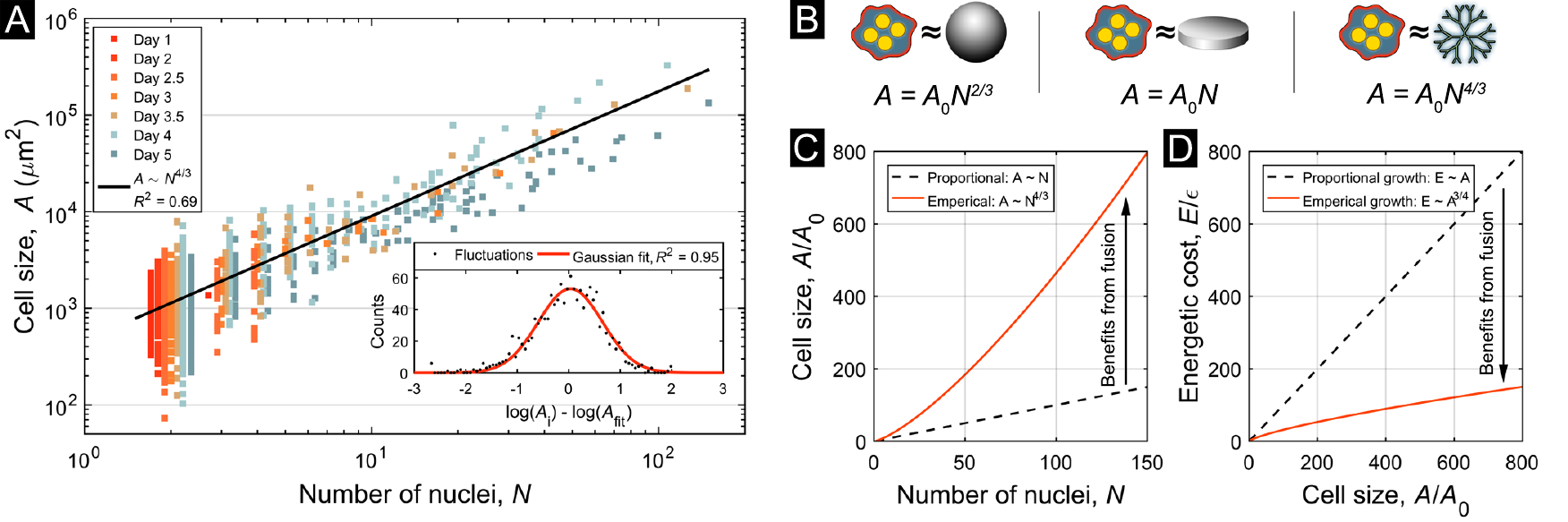}
\caption{Physiological scaling, models, and benefits of cell fusion. (A) Measurements of cell size and number of nuclei at different time points for 1,304 cells show power-law scaling. Low $N$ data points are horizontally offset for clarity.  Inset shows zero-mean Gaussian fluctuations around the best-fit line.  (B) Illustrations of the spherical, cylindrical, and fractal network models used to predict scaling between $A$ and $N$.  (C) Plot of increased cell size associated with fusion compared to a non-fusion scenario.  (D) Plot showing the decreased energetic cost of cellular growth arising from fusion compared to a collection of mononuclear cells that cover the same area.  From experiments, we set $h = 1 \ \mu$m.  }
\label{fig3}
\end{figure*}

{\it Fusion Benefits for Physiological Growth.} Because the size of osteoclasts and giant cells plays a role in biological function\cite{bar2007osteoclast,ishii2008osteoclast}, we used our Google Maps viewer to measure the projected area $A$ and number of nuclei $N$ for 1,304 multinucleated cells.  These measurements are pooled from seven independent replicates, each fixed at a different time point.  Empirically, the data follows a scaling law $A = A_0 N^{\alpha}$ ($R^2 = 0.69$; Fig.~\ref{fig3}A) with zero-mean Gaussian scatter ($R^2 = 0.95$; Fig.~\ref{fig3}A, inset).  Geometric arguments\cite{ishii2008osteoclast} modeling the shape of cells as smooth spheres or squat cylinders (Fig.~\ref{fig3}B) predict $\alpha = 2/3$ and 1, respectively, while a more detailed model that considers membrane ruffles\cite{milde2015multinucleated} predicts even lower values (SM).  Remarkably, we found $\alpha = 1.31 \pm 0.08$, which is larger than these expectations and well-approximated by the fraction 4/3 (SM).  This specific numerical value is interesting because it appears in biophysical theories predicting physiological scaling.  While discussion continues over the original data motivating these models\cite{dodds2001re,glazier2005beyond}, the basis for these theories focuses on energy production, distribution, and availability across all levels of biological organization\cite{west1997general,west2002allometric}.  In general, these theories suggest optimal resource distribution is achieved by fractal-like networks, and the specifics of these networks are critical size-regulating factors.  In cells, the relevant network is the spatial distribution of individual mitochondrion, where their dynamic organization gives rise to collective fractal-like properties in the mitochondria as a whole\cite{hoitzing2015function}.  Thus, while an abundance of mitochondria are often required for the substantial energy-demands of multinucleated cell types\cite{ch1931mitochondria,jimenez2013large}, this theoretical interpretation of $\alpha$ suggests the spatial organization of mitochondria plays a previously unrecognized size-regulating role.

\begin{figure*}
\includegraphics[scale=.9]{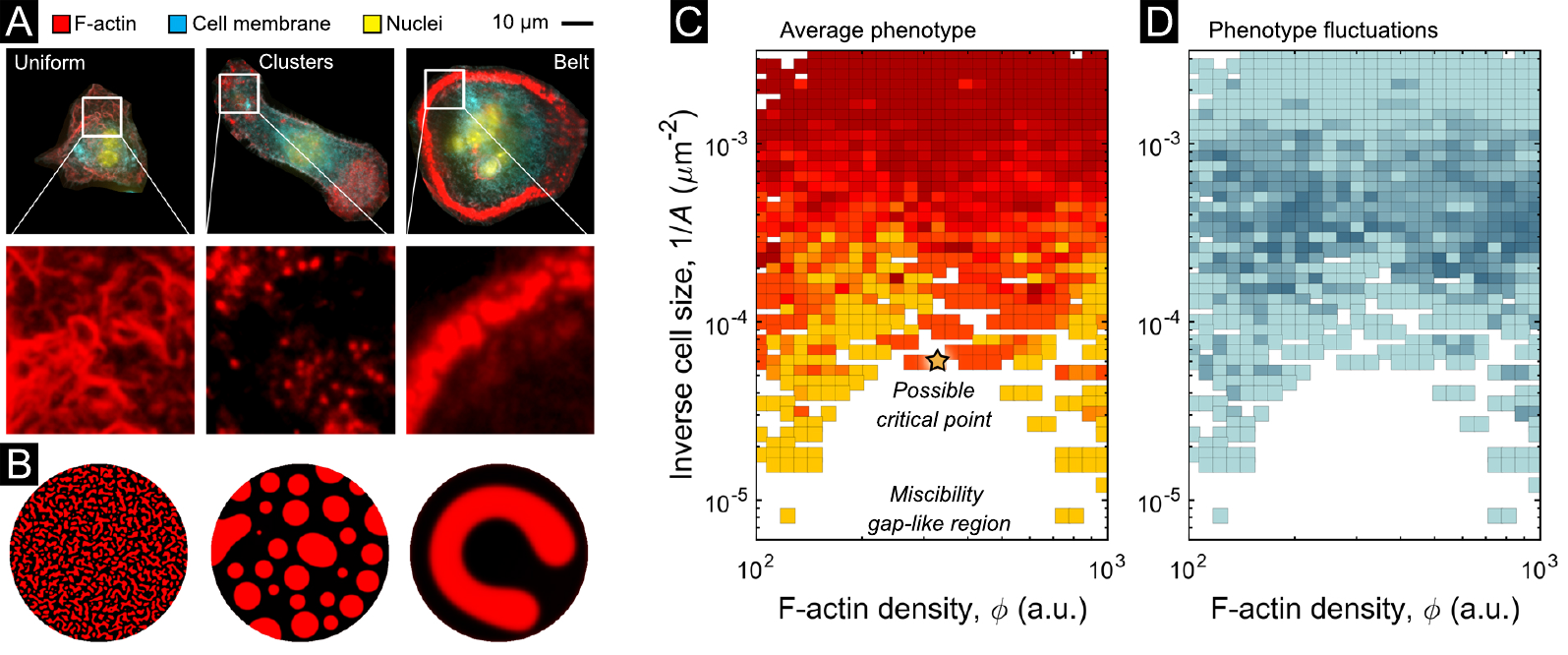}
\caption{Unsupervised machine learning suggests intracellular F-actin self-organizes by phase separation. (A) Classification by unsupervised machine learning finds F-actin organization linearly tracks with multinucleated cell development.  Zoomed boxes show typical F-actin phenotypes.  (B) Numerical simulations of the Cahn-Hilliard equation shows equilibrium morphologies from self-organized phase-separation resembles the F-actin organization in multinucleated cells.  These three simulations are generated by varying the interfacial energy relative to the thermal energy. (C) Phase diagram of F-actin organization shows phenotype distribution in phase space.  Inverse cell size $A^{-1}$ is the effective temperature; uniform (red), clustered (orange), and belt (yellow) phenotypes parallel thermodynamic phase diagrams.  Star is an estimate for the critical point.  (D) F-actin phenotype fluctuations are highest (dark blue) as cells transition from uniform to belt morphologies (upper and lower light-blue regions, respectively)}
\label{fig4}
\end{figure*}

While our data is ultimately agnostic to the origins of $\alpha$, values greater than 1 yield benefits-of-scale in function and efficiency.  For example, bone resorption by osteoclasts and foreign-body coverage by giant cells are biological functions of fused macrophages that rely on a large cell-substrate interface $A$.  To illustrate how fusion reduces the cost of covering an area $A$, we compare $N$ hypothetical cells of size $A_0$ to a single multinucleated cell formed from $N$ precursors of the same size.  While the $N$ cells cover an area $A = A_0 N$, the fused cells will grow to cover an area $N^{1/3}$ larger (Fig.~\ref{fig3}C).  In terms of consumed ATP/$\mu$m$^3$, unicellular eukaryotic and bacterial measurements show the metabolic cost of cellular growth generally dominates over the cost of maintenance\cite{lynch2015bioenergetic}.  As such, $N$ pancake-like mononuclear cells with thickness $h$ covering a total area $A$ have an energetic cost of growth $E = \epsilon A h$, where $\epsilon \approx 3 \times 10^{10}$ molecules of ATP/$\mu$m$^3$.  With fusion, the equivalent energetic cost of growth is instead $E = N \times (\epsilon A_0 h) = \epsilon A_0 h (A/A_0)^{3/4}$ (Fig.~\ref{fig3}D), where we exclude negligible costs of fusogenic protein assembly\cite{piques2009ribosome} (SM).  Therefore, compared to an equivalent number of mononuclear cells, an osteoclast with 15 nuclei has a $2.5\times$ increased size with 20\% energetic savings in growth, whereas a giant cell with 100 nuclei has a $4.6\times$ increased size and 68\% energetic savings in growth.  These calculations suggest an apparent benefit of fusion directly stemming from $\alpha > 1$: multinucleated cells can grow larger for a lower energetic cost, allowing the savings to be potentially reallocated to energy-demanding functions such as bone resorption and foreign-body rejection\cite{yagi2005dc, boissy2002transcriptional}.  While our argument generally applies for any example of fusion with $\alpha > 1$, increasing size can introduce practical or biological constraints not factored in this analysis\cite{galli2016cell, kyogoku2017large, lane2010energetics}, which when balanced against growth-promoting fusion, would ultimately define an optimum cell size in physiological contexts.

{\it Fusion Benefits for Self-Organization.} Classifying all 1,304 multinucleated cell images with unsupervised machine learning identified a linear progression of phenotype (SM; Figs.~S5-S7).  While $A$ and $N$ strongly correlated with phenotype classifications ($p < 10^{-47}$ and $10^{-66}$, respectively), F-actin organization was the principle feature associated with development.  The self-trained neural network found F-actin is uniformly distributed in small multinucleated cells, but as $N$ increases, F-actin organizes into small drop-like clusters, and eventually a dense belt (Fig.~\ref{fig4}A).  Remarkably, these classifications directly mirror the well-known stages of osteoclastogenesis, where F-actin reorganizes as the multinucleated cells mature in order to anchor the perimeter of osteoclasts during bone resorption\cite{jurdic2006podosome}.  Phenomenologically, these phenotypes also resemble condensation and phase separation in thermodynamic systems.  To explore this similarity, we numerically solved the Cahn-Hilliard phase separation equations in a circular domain (SM).  F-actin-like morphologies were easily reproduced, including the belt-like structure commonly seen in mature osteoclasts (Fig.~\ref{fig4}B, compare to Fig.~\ref{fig4}A top row).  With these observations in mind, we used fluorescence intensity to quantify F-actin density $\phi$, used the inverse cell size $A^{-1}$ as an effective temperature, and treated each cell as a probe of $\langle \phi, A^{-1} \rangle$ phase space (SM).  Pooling all the cells together allowed us to ask (i) whether a phase separation diagram shows evidence of a critical point and miscibility gap, (ii) whether cell phenotypes are consistent with what would be expected in each region of the phase diagram, and (iii) whether the fluctuations in phenotype increase near the hypothetical critical point and decrease at high/low effective temperature.  

Processing the image data led to a series of findings, each of which is consistent with a phase separation interpretation.  First, the empirical phase diagram contours suggest a critical point and miscibility gap (Fig.~\ref{fig4}C, star and empty region at low $A^{-1}$).  Second, the cell phenotypes are broadly distributed in the phase diagram in the same way as a conventional phase separation diagram: uniform F-actin distributions are at high effective temperature (Fig.~\ref{fig4}C, dark red), clustered F-actin appears as the temperature is lowered (Fig.~\ref{fig4}C, orange), and the F-actin belt is found below the critical point (Fig.~\ref{fig4}C, yellow).  Thus, as in conventional examples of phase separation, we see F-actin is uniformly distributed at high effective temperatures, but then separates into high-density and low-density domains as the effective temperature is decreased below a critical threshold.  Third, we observe the phenotype fluctuations are small at high effective temperature (Fig.~\ref{fig4}D, light blue at large $A^{-1}$), increase as the effective temperature approaches the critical point (Fig.~\ref{fig4}D, dark blue at intermediate $A^{-1}$), and decrease again as the effective temperature is further lowered (Fig.~\ref{fig4}D, light blue at small $A^{-1}$).  All of these observations are consistent with thermodynamic expectations, providing evidence of a phase separation occurring during osteoclastogenesis.  While multinucleated cells are definitively non-equilibrium, self-organization by phase transitions has the clear benefit of assembling ordered structures like the F-actin belt without explicit control over molecular placement\cite{tan2016self,su2016phase,kohler2011structure}.  

{\it Conclusion.} Having examined risks to population stability, benefits of supra-linear scaling in $A(N)$, and emergent phenotypes in multinucleated cells, we better appreciate the biophysical implications of fusion as a route to differentiation, specialization, and self-organization.  While some of these results are specific to macrophage-derived lineages, the risk-benefit framework and experimental tools introduced here are generalizable to study fusion in other developmental, biomedical, or industrial contexts.  Along these lines, we note the confocal fluorescence image data acquired during this study are not {\it prima facie} novel.  However, the scale of the data, which was enabled by new technological developments, is vastly larger than any previously reported.  Therefore, Big Data imaging directly impacted our ability to generate new quantitative interpretations of how cell-cell fusion relates to biophysical self-organization and development.  The core of the physical interpretations reported here are heavily inspired by approaches in non-living soft matter systems, and their continued application to other examples of cell-cell fusion will likely deepen our understanding of this common, yet still mysterious process.

\subsection*{Acknowledgments}
The authors thank T. Lambert and The Nikon Imaging Center at Harvard Medical School for the use of their microscopy facilities.  This work was funded by the National Institutes of Health under Awards 1R01EB018659-01, 1-U01-MH106011-01, AR062054, the National Science Foundation under Award CCF-1317291, and the Office of Naval Research under Awards N00014-13-1-0593, N00014-14-1-0610, N00014-16-1-2182, and N00014-16-1-2410.  JLS was supported by the National Cancer Institute of the National Institutes of Health under Award F32CA204038.  PY was supported by Wyss Institute Funds.  


\vfill \eject
\clearpage

\pagebreak
\begin{widetext}
\begin{center}
\textbf{\large Supplemental Materials: Population stability risks and biophysical benefits of cell-cell fusion in macrophage, osteoclast, and giant multinucleated cells}

\vspace{5mm}
Jesse L. Silverberg,$^{1,2}$ Pei Ying Ng,$^3$ Roland Baron,$^3$ and Peng Yin$^{1,2}$

\vspace{2mm}
$^1$\textit{Wyss Institute for Biologically Inspired Engineering, Harvard University, Boston MA 02115}
$^2$\textit{epartment of Systems Biology, Harvard Medical School, Boston, MA 02115}

$^3$\textit{Harvard School of Dental Medicine, Harvard University, Boston MA 02115}

\end{center}
\setcounter{equation}{0}
\setcounter{figure}{0}
\setcounter{table}{0}
\makeatletter
\renewcommand{\theequation}{S\arabic{equation}}
\renewcommand{\thefigure}{S\arabic{figure}}
\renewcommand{\thetable}{S\arabic{table}}

\section*{Mathematically modeling fusion-driven population dynamics}
To understand the consequences of fusion for population dynamics, we consider a simplified model where cell-cell fusion is treated in the same way as chemical reactions.  Using $n_i$ to denote the number of cells per unit area with $i$ nuclei, then the basic fusion reaction equation is $n_i + n_j \rightarrow n_{i+j}$.  The law of mass action says this symbolic representation can be written as the differential equation $d n_{i+j} / dt = k_{ij} n_i n_j$, where $t$ is time and $k_{ij} \equiv k_{ji}$ is the effective fusion rate specific to cells with $i$ and $j$ nuclei.  However, this expression is incomplete.  The rate of growth in $n_{i+j}$ also increases when cells with $i-1$ and $j+1$ nuclei fuse since $(i-1) + (j+1) = i + j$, which can be generalized to $i-k$ and $i+k$ for $i > k$.  Likewise, the rate of growth in $n_{i+j}$ decreases when cells with $i+j$ nuclei fuse with cells that have $\ell$ nuclei, $n_{i+j} + n_{\ell} \rightarrow n_{i+j+\ell}$.  As a concrete example, the full rate equation for $n_5$ is written by collecting all the terms for population growth and loss, $d n_5 / dt = k_{14} n_1 n_4 + k_{23} n_2 n_3 - \left( k_{15} n_1 n_5 + k_{25} n_2 n_5 + \ldots \right) = k_{14} n_1 n_4 + k_{23} n_2 n_3 - n_5 \sum_{\ell = 1}^{\infty} k_{5 \ell} n_{\ell}$.  Here, the positive terms consist of all $i$ and $j$ such that $i + j = 5$, while the negative terms consist of all $\ell \ge 1$.  More explicitly, the first few equations are:

\begin{eqnarray}
\frac{dn_1}{dt} & = & k_0 n_1 - n_1 \sum_{\ell = 1}^{\infty} k_{1 \ell} n_{\ell}, \nonumber \\
\frac{dn_2}{dt} & = & k_{11}n_1 n_1 - n_2 \sum_{\ell = 1}^{\infty} k_{2 \ell} n_{\ell},  \nonumber \\
\frac{dn_3}{dt} & = & k_{12}n_1 n_2 - n_3 \sum_{\ell = 1}^{\infty} k_{3 \ell} n_{\ell}, \nonumber \\
\frac{dn_4}{dt} & = & k_{13}n_1 n_3 + k_{22}n_2 n_2 - n_4 \sum_{\ell = 1}^{\infty} k_{4 \ell} n_{\ell}, \nonumber \\
\frac{dn_5}{dt} & = & k_{14}n_1 n_4 + k_{23}n_2 n_3 - n_5 \sum_{\ell = 1}^{\infty} k_{5 \ell} n_{\ell}, \nonumber \\
\frac{dn_6}{dt} & = & k_{15}n_1 n_5 + k_{24}n_2 n_4 + k_{33}n_3 n_3 - n_6 \sum_{\ell = 1}^{\infty} k_{6 \ell} n_{\ell}, \nonumber \\
& \vdots &
\label{model1}
\end{eqnarray}
With the assumptions here, there are no limitations on the maximum value $\ell$ can take, and the number of rate constants for fusion grows quadratically as ${\cal O}(\ell^2)$.  Thus, while this system of equations are relatively simple, the parameter space is quite large, making the dynamics challenging to treat analytically. 

We reiterate that, as stated in the main text, the model ignores details of biochemistry, apoptosis, genetic recycling, and regulatory signaling.  These simplifications allow us to narrowly isolate the effect of fusion in order to achieve a baseline understanding of its effect.  As a consequence, both apoptosis, which removes cells, and genetic recycling, which is a form of fission\cite{jacome2019developmental}, both contribute decay terms to Eqs.~(\ref{model1}) of the form $-\kappa_i n_i$, or simply renormalize the rate constants $k_{ij}$ to smaller values.  Given that cell death does not appear to be an issue at the time scale of the experiments and in the presence of nutrients, we safely set all $\kappa_i = 0$.  Furthermore, the renormalized rate constants from genetic recycling process is handled by our stochastic parameter sampling approach described in the next section, wherein we randomly select values for $k_{ij}$ between a physiological upper- and lower-bound.

\section*{Solving the fusion model with stochastic parameter sampling}
Solving Eqs.~(\ref{model1}) by stochastically sampling values for the fusion parameters sheds light on the general behavior of the model.  This computational method sets an initial population of precursor cells $n_1(t=0) = n_0$ with all other $n_i = 0$ for $i > 1$, and uses random values for all $k_{ij}$ and $k_0$.  The upper and lower bounds for the rates were fixed at once per hour and once per week, and were chosen from a uniform logarithmic distribution to obtain equal representation of all possible time scales between these bounds.  We coded the model in MATLAB, used the ODE45 integrator, and parallelized the computation on a 12-core CPU.  For each run with a randomly generated set of rate constants, we fixed $\ell_{\rm max}$, the maximum number of nuclei permitted, and $T$, the time-span to solve for.  Manually examining thousands of runs revealed the model's outcome was either a stable total population ($n_1(t)$ crashes) or a growing total population ($n_1(t)$ holds steady).  We automated this identification process by examining the values of $dn_1/dt$ and $dn/dt = d(\sum_i i \cdot n_i)/dt$ at the end of each run and classifying the results based on the relative rate of growth.  To generate statistics, we held $n_0$, $\ell_{\rm max}$, and $T$ fixed, then repeated the simulation $m$ times to determine what percentage of the randomly selected rate constants led to stable or growing populations (Fig.~S1A).  We generally fixed $n_0 = 150$ to reflect the experimental macrophage population density, fixed $\ell_{\rm max} = 8$ to maintain reasonable computation times, fixed $T = (60$ minutes/hour$) \times ( 24$ hours/day$) \times ( 5$ days$) = 7200$ minutes to emulate experiments, and fixed $m = 5000$ to average over fluctuations when determining the percentage of simulations that had stable or growing steady-state dynamics.  We calibrated the computational method by testing and confirming that the population outcomes were broadly independent from $\ell_{\rm max}$, $T$, and $m$.  Specifically, varying $\ell_{\rm max}$ showed the percentage of simulations leading to stable populations was $(80 \pm 1)\%$ for $\ell_{\rm max} = 5$, $(77 \pm 2)\%$ for $\ell_{\rm max} = 6$, $(80 \pm 2)\%$ for $\ell_{\rm max} = 7$, $(78 \pm 2)\%$ for $\ell_{\rm max} = 8$, $(80 \pm 2)\%$ for $\ell_{\rm max} = 9$, and $(80 \pm 2)\%$ for $\ell_{\rm max} = 10$.  Varying $T$ showed the percentage of simulations leading to stable populations changed from $(78 \pm 2)\%$ to $(68 \pm 1)\%$ when $T$ was increased 100-fold from 5 days to 500 days.  Varying $m$ showed the percentage of simulations leading to stable populations was $(77 \pm 7)\%$ for $m = 500$,  $(77 \pm 3)\%$ for $m = 1,000$, $(78 \pm 2)\%$ for $m = 5,000$, $(79 \pm 1)\%$ for $m = 10,000$, and  $(78 \pm 1)\%$ for $m = 50,000$.

\section*{Solving the fusion model with analytic simplifications}
By setting a maximum size for the number of nuclei in any given cell so that $\ell_{\rm max} = 3$, we can use an analytical phase portrait method to understand the model's dynamics.  While in-depth analysis has shown similar methods can be used to understand the subtle interplay between stability, resilience, and collapse\cite{dai2012generic, dai2015relation}, we restrict our attention to the binary outcomes identified with the stochastic parameter sampling method.  Simplifying Eqs.~(\ref{model1}) by dividing by $k_0$, renormalizing time such that $k_0 t \rightarrow t$, and using dot-notation where $\dot{n}_i = dn_i/dt$, leads to
\begin{eqnarray}
\dot{n}_1 & = & n_1 - \beta n_1^2 - \gamma n_1 n_2, \nonumber \\
\dot{n}_2 & = & \beta n_1^2 - \gamma n_1 n_2, \nonumber \\
\dot{n}_3 & = & \gamma n_1 n_2,
\label{model2}
\end{eqnarray}
where $\beta = k_{11}/k_0$ and $\gamma = k_{12}/k_0$.  The nullclines in the $\langle n_1, n_2\rangle$ projection are calculated by solving $\langle \dot{n}_1, \dot{n}_2 \rangle = \langle 0, 0 \rangle$, which leads to four solutions.  Two are given by the lines $n_2 = (\beta/\gamma) n_1$ and $n_2 = \gamma^{-1} - (\beta/\gamma) n_1$.  The other two are degenerate and given by $n_1 = 0$.  Furthermore, the $n_1 = 0$ line is a line of fixed-points corresponding to a population crash of precursor cells, while the intersection of the non-degenerate nullclines lead to a fixed-point $\langle \bar{n}_1, \bar{n}_2 \rangle = \langle (2\beta)^{-1}, (2\gamma)^{-1} \rangle$ corresponding to steady total population growth due to a constant value of the precursor cell density $n_1$.  We numerically integrated Eqs.~(\ref{model2}) with MATLAB's ODE45 integrator for $\beta$ between 0.2 and 100 and plotted the total nuclei density $n(t) = n_1(t) + 2\times n_2(t) + 3\times n_3(t)$ to confirm the transition between stable and growing total population (Fig.~S1B).  While these trajectories had the same starting point $\langle n_1/n_0, n_2/n_0, n_3/n_0 \rangle = \langle 1, 0, 0 \rangle$, the terminal point of these phase trajectories showed a transition from the stable precursor cell population at $\langle (2\beta)^{-1}, (2\gamma)^{-1} \rangle$ to the crashed precursor cell population at the $n_1 = 0$ line as the $\beta$ was varied (Fig.~S1B, inset).  These outcomes are robust with respect to the initial population $n_0$, and the relative fusion rates $\beta$ and $\gamma$ (Fig.~S2).  Moreover, the analysis method is consistent with the stochastic parameter sampling, and validates reasoning presented in the main text.  

Interestingly, we see that when $\beta = \gamma = 0$, Eqs.~(\ref{model2}) reduces to an exponential model for growth of $n_1$.  When $\beta \ne 0$ and $\gamma = 0$, Eqs.(\ref{model2}) reduces to the logistic model of growth for $n_1$ where the population is known to plateau to a carrying capacity that depends on the value of $\beta$.  Once the population of $n_1$ stabilizes in the $\gamma = 0$ limit, $n_2$ will steadily increase at a constant rate.  Thus, Eqs.~(\ref{model2}) with $\beta, \gamma \ne 0$, represent the next level of model complexity by allowing for cells with 3 nuclei.  The $-\gamma n_1 n_2$ terms affecting $\dot{n}_1$ and $\dot{n}_2$ drive differences from these more well-understood models and add necessary complexity to accommodate the delayed uptick and plateau in population density observed in the empirical data (Fig.~2C).

\section*{Sample preparation}
\paragraph{Reagents.} Wheat germ agglutinin (WGA) Alexa Fluor 488 conjugate, Rhodamine-conjugated phalloidin, and Hoechst 33258, pentahydrate (bis-benzimide) were all purchased from Molecular Probes by Life Technologies (Eugene, OR, USA). Recombinant mouse RANKL was purchased from R\&D Systems Inc. (Minneapolis, MN, USA). Cell culture media, GlutaMAX$^{\rm TM}$, and penicillin-streptomycin were purchased from Gibco by Life Technologies (Grand Island, NY, USA), whereas fetal bovine serum was purchased from Gemini Bio-Products (West Sacramento, CA, USA). 

\paragraph{Macrophages and osteoclast culture.} Primary bone marrow macrophages (BMM) were extracted from tibiae and femora of 6-8 week-old C57BL/6 mice, and cultured under 5\% CO2 at 37$^{\circ}$C in $\alpha$-MEM supplemented with 10\% FBS and CMG 14-12 cell culture supernatant as source of M-CSF as previously described\cite{takeshita2000identification}. For imaging purposes, primary mouse bone macrophages were cultured directly on polymer coverslip culture dishes ($\mu$-Dish 35 mm, high, ibiTreat, \#1.5 polymer coverslip, tissue culture treated, sterilized) (ibidi USA, Inc., Fitchburg, WI, USA), and stimulated with recombinant mouse RANKL (10 ng/ml) every 48 hours.  

As shown in the main text, this protocol led to multinucleated cells with vastly different areal sizes, approaching a $10^3$ to $10^4$-fold difference between the largest and smallest cell (Fig.~\ref{fig3}A).  While \textit{in vivo} growth is generally not as extreme, active processes such as bone resorption makes it difficult to isolate fusion's effect on cell size and development.  Nevertheless, the pro-growth conditions of diffuse and steady RANKL signaling used in this protocol allow us to isolate the effects of fusion on population dynamics.

\paragraph{TRAP staining.} Mouse BMM precursor cells were seeded at a density of $6 \times 10^3$ or $1 \times 10^4$ cells per well in a 96-well tissue culture plate, and stimulated with varying concentrations of RANKL (10 to 50 ng/ml) over a course of 4 days. At the end of Day 4, the cells were fixed using 4\% PFA and stained for TRAP activity using the Leukocyte Acid Phosphatase TRAP kit (Sigma-Aldrich, St Louis, MO, USA) according to kit instructions. Briefly, pre-warmed deionized water (37 C) was added to the following solutions provided in the kit: 1) napthol AS-BI phosphate solution, 2) acetate solution, 3) tartrate solution, and 4) diazotized Fast Garnet GBC solution. The solutions were mixed gently by inversion and added to the cells. This was followed by a 30 minute incubation at 37 C in the dark. At the end of the incubation period, the staining solution was removed and the cells were rinsed gently with $1\times$ PBS twice, before imaging with light microscopy (Fig.~S3).

\paragraph{Immunohistochemistry.} Macrophages and osteoclasts were fixed with 4\% paraformaldehyde in PBS for 15 minutes at room temperature, followed by incubation with WGA for 30 minutes in Hank's balanced salt solution before cell permeabilization for 5 minutes with 0.1\% Triton X-100 in PBS. F-actin was stained with Rhodamine-conjugated phalloidin and nuclei visualized with Hoechst 33258.

\section*{Microscopy}
All images were collected with a Yokogawa CSU-X1 spinning disk confocal on a Nikon Ti-E inverted microscope equipped with Plan Fluor 40x NA 0.75 objective lens.  Rhodamine phalloidin and WGA fluorescence was excited with, respectively, the 561 and 488 nm lines from a Spectral Applied Research LMM-5 laser merge module with AOTF controlled solid state lasers.  Emission signal was collected with a Chroma quad band pass dichroic mirror using 620/60 and 525/50 emission filters.  Hoechst fluorescence was excited using widefield illumination from a Lumencor SOLA and 395/25 excitation, 400lp dichroic and E420LP emission filters.  Confocal and widefield images were acquired with the same Hamamatsu ORCA-R2 cooled-CCD camera controlled with MetaMorph 7.8.13 software, using an exposure time of 150 ms.  

Tile scans with 20\% overlap in both the $x$ and $y$ directions between adjacent images were collected with a Prior ProScan III motorized stage.  At each time-point, 25 to 45 $z$-series optical sections were collected with a step-size of 1.0 $\mu$m using the Nikon Ti-E internal focus motor.  Even though multinucleated cells were generally uniformly flat, the large number of $z$-slices were required to account for slight tilting of the culture coverslips.  Images were saved as gray-scale multi-paged 16-bit $1344 \times 1024$ TIFF files with 0.162 nm $\times$ 0.162 nm size pixels.  Each page of the TIFF file corresponds to a single $z$-plane.

\section*{Image data processing with the Molecular Atlas Platform (MAP)}
The principle challenge for large-scale imaging experiments is in handling quantities of data that vastly exceed RAM capacity.  Of the algorithms and software emerging to tackle this problem, a common theme is strategic caching and memory management.  We therefore constructed a software package in MATLAB for (i) processing raw microscopy data into gigapixel-scale images, (ii) converting these images into an easy-to-view format compatible with the Google Maps JavaScript API, and (iii) quantification with cloud-based measurement/annotation/analysis tools.  In this section, we provide a detailed description of how the platform functions and addresses common challenges in big data imaging.

A typical consumer PC has between 8 and 32 GB of RAM, which is smaller than the typical imaging experiment performed in the main text by a factor of 10 to 20.  Therefore, we developed the Molecular Atlas Platform (MAP) as a MATLAB tool to address this mismatch between quantity of data generated and computational capacity.  First, MAP preallocates memory, typically around 2 to 5 GB, for a single 16-bit gigapixel-scale image.  Starting with the first $z$-plane, MAP selectively reads a single page from each TIFF image to reconstruct the stacks one plane at a time.  When the page is read, MAP applies flat-field\cite{model2014intensity, kask2016flat} and background corrections, then determines whether the tile is in a corner, edge, or centrally-located region based on the file metadata and known overall dimensions.  Depending on where the tile is located within the larger image, the 20\% overlap between tiles is used to linearly blend two, three, or four sides with its neighbors. This process reduces artifacts from drift and optical distortion.  After the $z$-plane is reconstructed from individual tiles, it is written to long-term storage and the preallocated RAM reused for the next $z$-plane.  This process repeats for all $z$-planes, and then for all acquisition channels.  Once completed, preallocated RAM is released and made available for further processing.

Next, MAP uses the MIJ Java package to interface MATLAB with ImageJ.  Using a scripted sequence of commands, MAP instructs ImageJ to load the stitched $z$-planes as a virtual stack and perform a max projection.  This projection is saved at full 16-bit depth with lossless LZW TIFF compression to preserve the quantitative relationship between pixel intensity and fluorescence.  However, these images are often difficult to compare between channels due to differences in dynamic signal range.  Therefore, MAP instructs ImageJ to perform an auto brightness/contrast adjustment, resample the image to 8-bit depth, and save with lossless LZW TIFF compression to a second file.  Saving the data twice using different parameters allows us to maintain quantitative rigor for downstream analysis with the first data set, while also being able to present the data in the aesthetically clearest possible way with the second data set.  

Once MIJ has completed all $z$-stack compressions, MAP uses the 8-bit projected grayscale images for color composition and layer merging.  For compatibility with the web-based Google Maps JavaScript API, this step of the MAP pipeline must process output for all possible color combinations.  As an example with 3 colors symbolically denoted as A, B, and C, the software must create colorized images corresponding to the color combinations: A, B, C, AB, AC, BC, and ABC.  This processes starts by selecting one of the 8-bit contrast-adjusted grayscale projected images and preallocates RAM for an 8-bit RGB image with the same dimensions.  The grayscale image is then colorized with a user-specified color using scalar multiplication of the RGB triplet and the grayscale value.  When generating single-channel images, this colorized image stays in RAM and is passed to the next stage.  If multiple channels are being combined, then the next grayscale layer is loaded into memory, and in a single step, colorized and combined with the previous image using a screen blend.  Screen blends are symmetric with respect to the top and bottom image being combined and are computed according to $f(A,B) = \mathbbm{1} - (\mathbbm{1} - A)(\mathbbm{1} - B),$ where the output $f(A,B) = f(B,A)$ is the combined image.  Moreover, for multiple colors being combined, screen blends have the further symmetry property that $f(A, f(B,C)) = \mathbbm{1} - (\mathbbm{1} - A)(\mathbbm{1} - B)(\mathbbm{1} - C)$, which can be generalized for an arbitrary number of layers.  A critical challenge in this computation is that the RAM has allocated memory for 8-bit variables, but certain combinations of intensity values can lead to values that exceed the range of an 8-bit variable during intermediate calculations of $f(A,B)$.  Thus, when computing $f(A,B)$, MAP temporarily allocates additional memory for the intermediate outputs that exceed 8-bit capacity.  After all colorization and composition steps are completed, the image stays in RAM and is passed to the next stage.

To be compatible with the Google Maps JavaScript API, MAP takes the current image in memory, regardless of what combination of layers or colors it represents, and recursively scales and slices the image into tiles that are $256 \times 256$ pixels in size.  At the lowest scale, the entire image is fit into a single $256 \times 256$ pixel tile.  At the next scale, the image is fit into four $4 = 2 \times 2$ tiles for a total $512 \times 512$ pixel image.  At the third scale, the image is fit into $16 = 4 \times 4$ tiles, etc.  The individual tiles are saved as high-quality JPG images that are around 1 to 2 kb per tile, and with a naming scheme that indicates both position and zoom level.  A single image with $z_m$ zoom levels leads to $\sum_{z = 0}^{z_m} 4^{z}$ total tiles.  For 10 zoom levels, this translates to 1,398,101 small JPG files.  In a 3 color experiment with 7 possible combinations of channels, MAP then generates 9,786,707 files from the gigapixel-scale imagery.

In addition to outputting scaled and sliced image tiles, MAP also generates a web-ready HTML document that combines custom JavaScript, JSON, and CSS code to create an interface between the Google Maps JavaScript API and the resliced colorized image data.  The HTML document is a User Interface (UI) that can be viewed locally on a PC, or online from any internet-connected computer when the sliced images and HTML document are uploaded to a server.  In the same way Google Maps uses overlays to mix-and-match illustrated and satellite imagery, we repurposed the same layering functions to display all combinations of the 3 channels in our microscopy experiment.  Tools for optionally selecting which layers to view at any given moment were coded into the UI, which allows for instantaneous flipping back-and-forth between the possible single-color and multi-color views.  In addition to functioning as a platform for viewing experimental microscopy data, we also developed a suite of analysis tools as part of the UI.  These tools use KML data layering to allow the user to (i) add pins to mark regions of interest, (ii) measure distances between two points on-screen with a dynamically adjustable pre-calibrated ruler, (iii) select specific regions of interest for quantitative analysis in MATLAB or 3D rendering, and (iv) draw editable shapes including lines, arbitrarily sized $n$-gons, and squares.  All of these annotation tools were designed so they could be saved in the cloud in a Google Fusion Table for later recall by another user or for another analysis session on any internet-connected desktop, laptop, or mobile device.  

For the experiments performed here, raw microscopy data was processed through the entire MAP pipeline and the resulting JPG and HTML files were uploaded to a local web server (Figs.~2, 3, and S4).  Cell nuclei were manually identified and marked using the pin annotation tool (Fig. 2B).  This process involved (i) selecting the tool from the menu and (ii) clicking on individual nuclei.  Each time the user clicked, a pin was added to the field of view at the place where the click occurred.  This pin's position was adjustable at the time it was added to allow for corrections, but ``locked'' once the data was saved to the cloud to ensure research integrity.  The pin tool generates $(x, y)$ geospatial coordinate data for each pin added by the UI.  We saved this coordinate data for later analysis.  Multinucleated cells were identified by visual inspection and their perimeter was outlined with the $n$-gon drawing tool.  This process involved (i) selecting the tool from the menu and (ii) clicking along the outline of cells of interest.  This sequence of clicks defined a closed polygon that approximates the cell perimeter.  A greater number of clicks generally lead to a greater resolution of the cell's shape.  Across 10 measurements, we found a $< 5\%$ error in area $\delta A$ when comparing typical clicked boundaries to high-fidelity boundaries acquired with a 20-fold greater resolution in the perimeter.  This difference in area is generally small, and consistent with the fact that errors in perimeter $\delta r$ are of second-order smallness when measuring area according to $\delta A / A \sim (\delta r / r)^2$.  Consequently, larger cells have a more precise measurement of $A$, indicating that the large values of area critical for determining the exponent $\alpha$ have a high degree of precision, which helps compensate for the greater scarcity of data on cells with $N \ge 50$ (Fig.~3A).  As with the pin tool, each $n$-gon is adjustable at the time it is added, but ``locked'' once the data is saved to the cloud to ensure research integrity.  For the purposes of our analysis, the level of accuracy we required from the pin tool was minimal.  To count the number of nuclei $N$, we simply require the pin tool's $(x,y)$ position be within the perimeter of the $n$-gon.  This is both easy to achieve and verify when performing the data annotations.  In MATLAB, the nuclei and $n$-gon coordinate positions were used to determine how many nuclei $N$ were within each multinucleated cell by algorithmically counting the number of nuclei within each $n$-gon's boundary.  No recorded nuclei were found outside the perimeter of a cell.  A scaling coefficient was used to convert from Google Map's geospatial coordinates to calibrated length dimensions measured in microns, enabling us to quantify the cell areal size $A$ in $\mu$m$^2$ from the $n$-gon perimeter.  Perimeter coordinates were also used to extract fluorescence image data for each multinucleated cell from the pre-contrast-corrected max-projection images.  This allowed us to algorithmically isolate each cell's fluorescence intensity and make quantitative measurements of the F-actin density, $\phi$.

\section*{Why have two pipelines when saving MAP data?}
In describing MAP, we stated it is generally ``difficult to compare between [colorized image] channels due to differences in dynamic signal range.''  The challenges of dynamic signal range in a multicolor image is a standard problem when performing quantitative image analysis on microscopy data.  As such, there is also a standard solution: having two datasets saved in the manner as-described in the Methods enables us to overcome limitations imposed by dynamic signal ranges.  This dual track approach simultaneously preserves \textit{both} high-quality image visualization \textit{and} quantitatively-reliable image data.  For clarity, the challenges imposed by dynamic signal range occurs in the following hypothetical scenario:

\begin{enumerate}
\item A two-color 16-bit image has fluorescent target A recorded with pixel intensities varying from 250 to 1,000, and fluorescent target B recorded with pixel intensities from 250,000 to 1,000,000.   The range between the upper and lower bounds is called the ``dynamic signal range,'' and in the example here, the dynamic ranges are quite different - a range of 750 for A, and a range of 750,000 for B.  

\item If target A is assigned to the red channel and target B is assigned to the blue channel, the two channels can be merged to a red-blue composite image.  Because images are generally displayed on an 8-bit 0 to 255 RGB scale, the goal is to map measured signal intensities (pixel values) to colors in an RGB-compatible fashion.  The difficulty arises when we ask \textit{how} to best do this mapping.  

\item One approach would be to set 0 to black and 1,000,000 to the maximum color saturation (e.g., 255 in both 8-bit red and blue channels).  As a result, the composite image would be fully dominated by blue because the pixel intensity values are 1,000-fold higher.  While our eyes would completely miss the red signal (red's strongest value is 1/250$^{\rm th}$ the magnitude of the weakest blue signal), the image would be quantitatively precise in terms of the ratio of fluorescence signals.  This outcome would be desirable for quantitative analysis, but practically useless for visually interpreting the image data.

\item An alternative approach is to scale the dynamic ranges on a per-channel basis.  In this case, the red channel would set 0 to black and 1,000 to pure red, and the blue channel would set 0 to black and 1,000,000 to pure blue.  The composite would now show a mixture of red, blue, and various shades of purple because the dynamic signal ranges were independently scaled.  However, this image would be inappropriate for quantitative analysis, as any measurement of relative red-to-blue signal would be off by a factor of 1,000.  Nevertheless, it would be easily to visually interpret.
\end{enumerate}

The scenarios just described raises a further question: What is the most appropriate treatment for composing multicolor images when handling digital image data?  The answer, as is often the case, depends on the application.  If the researcher wants a quantitative measure of relative concentrations as determined by fluorescence signal, than the first approach is better.  If the research question simply involves detecting a co-localization or a binary yes/no answer, than the second approach is fine. 

As MAP is a general purpose tool for multiscale imaging, we make no presumptions about the user's specific needs and therefore impose no irreversible restrictions during the analysis pipeline.  Returning to the main question of this section -- why have two pipelines when saving MAP data -- the answer is simple.  One dataset is retained for quantitative analysis (e.g., the first option in our hypothetical 2-color scenario), while the second dataset is used for generating images that can be visually interpreted (e.g., the second option in our hypothetical 2-color scenario).  Without dual tracks for saving output, we could end up in a situation where either (i) the differences in dynamic range would cause a single fluorescence signal to dominate the image, or (ii) any quantitative measurements are incorrectly reported.  Because hard drive space is inexpensive, it's completely reasonable to save two sets of data to have both options -- high quality visualization \textit{and} quantitatively reliable measurements -- simultaneously available.

\section*{Validating the fusion model with experimental data}

Having introduced an ODE rate-equation approach to studying cell-cell fusion population dynamics, we now seek to validate the solutions to Eqs.~(\ref{model1}) and (\ref{model2}).  One potential method is to perform an experiment with RANKL titration to test whether this fusion-initiating ligand allows for a systematic probing of the model's phase diagram (Figs.~S1 and S2).  However, this hypothetical experiment assumes a simple linear relationship between the RANKL concentration and the effective fusion behavior.  In fact, a RANKL dose that's too small fails to initiate fusion, while increasing the RANKL dosage beyond a point no longer increases the rate of fusion\cite{rahman2015proliferation}.  This nonlinear relationship arises from the RANKL signaling pathway mechanics and makes a titration-based assay both difficult to analyze and inconclusive in terms of validating the model.  Instead, we can accomplish this validation by comparing the measured and predicted time-dependent density of nuclei (Fig.~2B).  In addition, the same approach allows us to compare these data to conventional exponential and logistic models of growth.  

Specifically, the ODE and solution for exponential population growth when only mononuclear cells are considered is
\begin{equation}
\frac{d n_1}{dt} = k_0 n_1 \quad \rightarrow \quad n_1(t) = n_0 e^{k_0 t},
\label{model3}
\end{equation}
where $n_0$ is the initial density of cells per unit area at $t = 0$, and $k_0$ is the growth rate.  Fitting Eq.~(\ref{model3}) to the empirical data (Fig.~2B) shows $k_0 = (0.46 \pm 0.27)$ (day)$^{-1}$.  

The logistic model advances exponential growth by allowing for a finite carrying-capacity $\kappa$ to exist within the system.  Thus, the relevant ODE and solution are 
\begin{eqnarray}
\frac{d n_1}{dt} = k_0 n_1 - \kappa n_1^2 \quad & \rightarrow &  \nonumber \\
\quad n_1(t) & = & \frac{n_0 e^{k_0 t}}{1 + n_0 \left(\frac{\kappa}{k_0}\right) \Big( e^{k_0 t} - 1 \Big)}, \nonumber \\
\label{model4}
\end{eqnarray}
where $n_0$ is again the initial density of cells per unit area at $t = 0$, and $k_0$ is the growth rate.  The best-fit to our data shows $k_0 = (0.56 \pm 0.36)$ (day)$^{-1}$, and $\kappa = (3.3 \pm 9.8)\times 10^{3}$ (area / day).  Note that the large 95\% confidence interval on $\kappa$ indicates a wide range of values are permissible resulting in a similar $R^2$ with the optimal value centered on $3.3 \times 10^3$ (area / day).  From this observation of the fitting results, we can conclude that $\kappa$ does not have a strong influence on the quality of the fit and the shape of the curve is insensitive to values in this range.  Consequently, the failure of the logistic model to accurately capture the empirical population dynamics trend is robust.  

For completeness, the best-fit values and 95\% confidence intervals for the fusion model are $k_0 = (4.1 \pm 20)$ (day)$^{-1}$, $k_{11} = (10 \pm 150) \times 10^{-3}$ (area / day), and $k_{12} = (7.9 \pm 12.7) \times 10^{-3}$ (area / day).  Here again, we see the rate constants have large confidence intervals indicating a range of values lead to similarly high-quality best-fits.  While model fitting is often used to determine specific numerical values for model parameters, we stress that in our case here, we have used model fitting to compare the performance of multiple competing models, and \textit{not} to determine precise values of the parameters themselves.  We can therefore conclude that while a range of parameter values are permissible within each model, exponential and logistic growth do not have the necessary complexity to recapitulate the observed population dynamics.  The cell fusion model in Eqs.~(\ref{model1}) and (\ref{model2}), however, does have the requisite complexity to capture the empirical data, and thus offers a measure of validation.  Nevertheless, the complexity that allows us to validate the model is precisely the same complexity that leads to the potential for catastrophic population crashes, as discussed in the main text (Fig.~1D).

\section*{Geometric models for $\alpha$}
In the main text we considered two simple geometries with smooth surfaces to understand the relationship between the areal size $A$ and number of nuclei $N$.  Here, we show how to derive those relationships.  Treating mononuclear precursor cells as spheres with radius $r_0$ and assuming all cells have a volume $V_0 = (4/3) \pi r_0^3$, then each mononuclear precursor cell will also have a projected area $A_0 = \pi r_0^2$.  When $N$ of these spherical cells fuse, the total volume increases to $N \times V_0 = (4/3) \pi N r_0^3 = (4/3) \pi r_N^3$.  From this equality, we see the multinucleated cell has a radius $r_N = N^{1/3} r_0$, and therefore a projected area $A = \pi r_N^2 = \pi r_0^2 N^{2/3} = A_0 N^{2/3}$.  Hence, $\alpha = 2/3$ as stated in the main text.  Treating mononuclear precursor cells as stout cylinders with radius $r_0$, height $h$, and taking a similar approach, we instead have a volume $V_0 = \pi r_0^2 h$ with projected area $A_0 = \pi r_0^2$.  Fusing $N$ of these precursors together leads to a volume $N \times V = \pi N r_0^2 h = \pi r_N^2 h$, from which we find $r_N = N^{1/2} r_0$.  Thus, assuming $h$ remains fixed, the projected area of a multinucleated cylindrical cell is $A = \pi r_N^2 = \pi r_0^2 N = A_0 N$, leading to $\alpha = 1$.  

To validate our results and test for false-positives, we performed constrained fits where we fixed $\alpha$ and optimized for the proportionality constant $A_0$ in the model equation $A =  A_0 N^{\alpha}$.  When we fixed $\alpha = 2/3$ for the cell-as-spheres model, we found the best-fit $R^2 = 0.539$ and the data points scattered about this line with a Gaussian distribution whose quality of fit was $R^2 = 0.95$ and center was offset by $4.7$ standard deviations from zero.  This strong offset indicates the model has systematic deviation from the data and therefore poorly describes the experimental observations.  Likewise, when we fixed $\alpha = 1$ for the cell-as-cylinders model, we found the best-fit $R^2 = 0.66$ and the data points similarly scattered with a Gaussian distribution.  In this case, the distribution's $R^2 = 0.93$ and its center was offset by 5.3 standard deviations from zero, again indicating strong systematic bias disfavoring the model.  In the best-fit model where $\alpha = 1.31 \pm 0.08 \approx 4/3$, we see the Gaussian scatter (Fig.~3A, inset) is offset from zero by 0.037 standard deviations.  As such, these spherical and cylindrical models fail to capture the empirical trends.

As a modification to simple smooth geometries, we consider the scaling between $A$ and $N$ that results from a wrinkled surface to accommodate the effects of extensive membrane ruffles observed in the osteoclast cell type\cite{milde2015multinucleated}.  Qualitatively, the true area $a$ will be larger than the apparent area $A$ due to extra surface hidden in the membrane invaginations.  As a result, we anticipate the value for $\alpha$ to be decreased by wrinkling when calculating the relationship between the observed area and the number of nuclei.  To quantitatively demonstrate this expectation, we consider a simple spherical geometry with apparent surface area $A$ covered in wrinkles at a density of $n$ wrinkles per unit area.  Thus, the total number of wrinkles on the surface $nA$ multiplied by the typical wrinkle amplitude $u$ and typical wrinkle wavelength $\lambda$ yields an estimate $a \approx (nA)(u\lambda)$, which is the true total surface area.  Wrinkles on surfaces and biological membranes are well-known to be patterned by mechanical instabilities that arise from a strain $\varepsilon$ that exceeds a critical value $\varepsilon_c$.  This strain generally comes from an excess surface area for a given volume leading to a mechanically-selected balance between bending and stretching deformations of the surface\cite{landau1959course, audoly2010elasticity, stoop2015curvature}.  Linear stability analysis shows this balance selects a wrinkling wavelength $\lambda$ that depends on the membrane thickness and mechanical properties of the cell including the Young's moduli.  Importantly, the wrinkling wavelength is insensitive to the size of the cell when patterning occurs.  The wrinkling amplitude, however, is size-dependent because the strain $\varepsilon$ depends on the difference between apparent surface area $A$ and true surface area $a$.  This wrinkling amplitude scales according to $u \propto (\varepsilon/\varepsilon_c - 1)^{1/2}$.  As the cell grows, we have $\varepsilon \sim \sqrt{a} - \varepsilon_c$ so that $u \sim a^{1/4}$.  Because only the amplitude depends on cell size, the density of wrinkles $n$ remains steady and we have $a \propto A u \propto A a^{1/4} \rightarrow a^{3/4} \propto A$.  As for the smooth geometries considered above, volume conservation relates the number of nuclei $N$ to the true surface area by $a \sim N^{2/3}$, which upon substitution into the expression for the apparent surface area, yields $A \sim (N^{2/3})^{3/4} = N^{1/2}$.  Thus, in the case of wrinkled spheres $\alpha$ is reduced from $2/3$ to $1/2$ by the surface area hidden in invaginations.  If this more detailed approach to modeling the observed scaling between $A$ and $N$ is any indication, then perhaps the measured value of $\alpha$ is even more unusual than the simple geometric models in the main text suggests.

\section*{Allometric model for $\alpha$}
As referenced in the main text, there has been substantial theoretical effort focused on understanding physiological scaling relations between body mass $M$ and metabolic rate $B$.  Rather than re-derive these relationships here, we show how those existing results could relate to ours.  In particular, we start with $B = B_0 M^{3/4}$, which has been the central empirical claim that theoretical models have attempted to account for.  In this expression $B_0$ is the scaling relation's coefficient and is generally expressed in terms of the model's microscopic parameters.  For our purposes, we simply note that body mass $M$ is proportional to the volume $V$ by the density $\rho$.  In the case of multinucleated cells produced by macrophage fusion, we find they are typically thin and have uniform height $h$, so that to a reasonable approximation $M = \rho A h$.  Metabolic activity $B$, on the other hand, is related by the energetic output of the mitochondria.  If each mononuclear precursor cell contributes an equal amount of mitochondrion to the mitochondria, then there will be a proportionality between the metabolic output and the number of nuclei.  Thus, we hypothesize $B = b N$.  Substituting the expressions for $M$ and $B$ into the scaling relation yields $b N = B_0 (\rho A h)^{3/4}$, which can be rearranged as $A = (\rho h)^{-1} (b/B_0)^{4/3} N^{4/3} = A_0 N^{4/3}$.  This calculation is intended to be an illustrative example of how the empirical scaling can arise in multinucleated cells rather than a concrete proof that optimized energy distribution through fractal-like networks is the causal mechanism for the data (Fig.~3A).  Nevertheless, experiments investigating ATP trafficking could be used to determine whether these optimized transport models provide useful mechanisms to interpret internal organization in multinucleate cells (Fig.~3B).  Validation of this interpretation would also likely involve direct measurements of the diffusive properties of the mitochondria to show non-Brownian behavior, as well as specific signatures in the distribution of distances between individual mitochondrion.

\section*{Back-of-the-envelope estimate for energetic costs of fusogenic protein assembly}
When comparing the energetic differences between cell growth and cell fusion (Fig.~3D), we ignore the ATP consumption involved in generating and operating fusogenic machinery.  Because fusogenic proteins typically convert a receptor-ligand bond into mechanical strain on the membrane, we ignore the energetic costs of operating these proteins and instead focus our attention on the costs of protein assembly.  In the specific context of macrophage fusion, we note the specialized proteins CD47 and Membrane Fusion Receptor (MFR) have the necessary Ig domains that function as a receptor-ligand pair between two cells (\textit{3}).  Respectively, they contain 323 and 504 amino acids, which at a cost of 5 ATP/amino acid\cite{piques2009ribosome}, require 1,615 and 2,520 ATP/molecule to synthesize.  For a typical precursor cell of projected size $A \approx 10^3 \ \mu$m$^2$, we can estimate a spherical surface area $\approx 4 \times 10^3 \ \mu$m$^2$.  If we assume both of these $\approx 5$ nm proteins are generally spaced $\approx 100$~nm apart and uniformly cover the entire cell surface, then we calculate $4 \times 10^5$ of each fusogen per cell for a total of $\approx 3 \times 10^9$ molecules of ATP required in protein assembly.  This estimate is $10$ times smaller than the cost of growth described in the main text, and we therefore neglect its contribution.

\section*{Deconvolving time-dependent measurements with quantitative image analysis and unsupervised machine learning}
In principle, plotting $A$ and $N$ as a function of time is expected to reveal the developmental trends of multinucleated cells (Fig.~S5A-D).  Instead, these data show a high degree of scatter (Fig.~S5A and C) with a skewed distribution at each time point.  We therefore generated a series of box plots to visualize a summary of the statistical trends where the top and bottom of each box, otherwise known as the interquartile range, are the 25$^{\rm th}$ and 75$^{\rm th}$ percentiles, the line in the middle of the box is the median, and the whiskers extend $1.5\times$ the interquartile range away from the box edges.  Data beyond the whiskers are considered outliers (e.g., Fig.~S5B, red `$+$' symbols).   Notches near the median line are computed to easily compare multiple time points for statistical significance.  Specifically, when two notches do not horizontally overlap (e.g., Fig.~S5B, day 1 and 4), the medians are different at the 5\% significance level.  Conversely, when notches do overlap there is no statistical significance.  This calculation implicitly assumes the data is normally distributed, however, significance comparisons with medians from non-Gaussian distributions is a generally robust procedure.  While a qualitative trend for increasing $A$ and $N$ is clear (Fig.~S5B and D), this quantitative analysis shows scatter suppresses statistical significance.

Evidently, a conventional method to analyze time-dependent data fails to adequately capture developmental trends (Fig.~S5E).  Largely, this failure arises because cell fusion is a massively parallel process with multinucleated cells regularly forming, which broadens and scatters the distributions for $A$ and $N$ over time (Fig.~S5F).  As an asynchronous process, fusion effectively convolves different developmental stages at a single time point resulting in the observed scatter of population-scale measurements (Fig.~S5A-D).  Thus, time becomes less effective at linearly tracking development.  In essence, this means time is the wrong parameter for understanding population-scale development.  

To better understand the development of multinucleated cells, we turned to machine learning and used a self-organizing map (SOM) neural network (NN) to cluster the multinucleated cell images into classes.  A key feature of SOMs is that the resulting classifications retain topological information about which classes are similar with others, and as such, can be used to recover the linear developmental progression.  Pixel coordinate data generated in the web-based Google Maps viewer and stored in a Google Fusion Table was downloaded from the cloud and imported into MATLAB for analysis.  These coordinates were used to extract individual cells from the gigapixel-scale images and independently save 1,304 pre-contrast-corrected images representing each multinucleated cell.  Quantitative image analysis was performed on each fluorescence channel measuring both the number of features using MATLAB's SURF feature detection algorithm as well as the histogram of all 3 fluorescence signals.  To remove interference from cells adjacent to the cell being analyzed, we used the coordinates of the multinucleated cell's perimeter as a mask to exclude extraneous signals.  We combined the feature detection hits and fluorescence intensity distributions with metadata including the (i) fixation time point, (ii) number of nuclei, (iii) cell areal size, (iv) cell perimeter, (v) cell roundness as measured by the ratio of the perimeter squared to the cell area, (vi) radial distribution of F-actin density, and (vii) radial distribution of nuclei.  All of this information was passed into an unsupervised machine learning algorithm using MATLAB's machine learning toolbox.  Training of the NN was performed with batch weighted/bias rules and performance was gauged with the mean squared error.  We tested NNs generated by MATLAB's \textit{selforgmap} function ranging in size from $2 \times 2$ to $9 \times 9$ with hexagonal topologies and varied the number of iterations from 100 to 100,000 (Fig.~S6A).  Broadly, the fully trained NNs identified three groups of cells (Fig.~S6B), though in the larger networks they presented as clustered neurons with well-separated distances.  These groups track closely with the known development of osteoclasts where F-actin organization transitions from a uniform, to clustered, and eventually to a belt-like phenotype.  

Because the boundaries of NN clusters that define groups depend on a random initial seed used in SOM optimization, we performed an adjacency analysis to determine the robustness of the method.  Specifically, we generated 50 randomly sized SOMs NN from $2 \times 2$ to $6 \times 6$, performed training for 5,000 iterations, and extracted the identity of the neuron each cell was assigned to.  In small NNs, a neuron could be associated with 900 or more cells, while in larger NNs, some neurons only represent a single cell.  Averaging over all 50 runs, we formed an adjacency matrix to determine how frequently each pair of cells were assigned to the same neuron.  We then plotted this adjacency matrix as a schemaball (Fig.~S6C), where each cell is a node on the outer ring, and each line connects cells that share a neuron in 85\% or more of the SOMs.  Because the vast majority of lines connecting cells are within their own group (Fig.~S6C, groups defined by red, orange, and yellow arcs), the density of connections appears as a nearly solid band (e.g., Fig.~S6C, dense bands near black perimeter).  This analysis suggests the three groups defined in our initial SOM NNs are quite robust and lends credibility to their identification.  There are, however, a handful of lines connecting cells from different groups (Fig.~S6C, inset; black $x$'s).  These cells represent $\approx 4\%$ of the overall data set and have phenotypes suggesting misclassified groupings.  Nevertheless, by identifying the similarities between cells, the machine learning method was able to deconvolve multinucleated cell development from the measurements of $A$ and $N$ at different fixation time-points.  Sorting $A$ and $N$ according to these groups shows strong statistical significance (Fig.~S7) not found in the original time-dependent analysis.

\section*{Cahn-Hillard phase separation}
The Cahn-Hilliard equation is derived from a free energy functional describing the spatial and temporal dynamics of a binary liquid-like system\cite{cahn1958free, eyre1998unconditionally}.  Classically, the components have an energetic penalty for mixing that produces a rich variety of self-organized morphologies.  The mixture concentration $c$ is governed by the partial differential equation $\partial c / \partial t = \nabla^2[\gamma^2 c^3 - c - \nabla^2 c]$, where time and spatial derivatives are written here in dimensionless length and time scales, $\gamma$ is the strength of the energetic penalty between phase boundaries relative to thermal energy, and $c = \pm 1$ corresponds to a pure phase.  Numerical integration of the Cahn-Hilliard equation was performed using Eyre's linearly stabilized integration scheme in MATLAB\cite{eyre1998unconditionally}.  We used a $512 \times 512$ meshed 2D square grid, then superimposed a circular boundary with diameter equal to the width of the entire mesh.  This setup defined an idealized round cell and no-flux boundary conditions that maintain a fixed total concentration of $c$.  Seeding the grid with random initial values for $c$ distributed so the total integrated value was $\int c \ dA = 0.3$, models a binary system with more of one substance than the other.  The equations were integrated until the morphology stabilized into an equilibrium configuration.  When the penalty $\gamma$ is low, the stable morphology is a filamentous network of discrete domains (Fig.~4B, left).  Increasing $\gamma$ and again numerically integrating the equations leads to a morphology where round droplet-like domains coalesce during the time-dependent transient.  As the system stabilizes, a number of droplets remain well-defined and well-dispersed throughout the circular domain (Fig.~4B, middle).  Increasing $\gamma$ further and letting the equations stabilize leads to a single highly-concentrated domain that tends to wrap around the perimeter (Fig.~4B, right).

\section*{Generating the F-actin phase diagram}
In the MAP data processing pipeline, a max-projection gigapizel-scale image was saved for each fluorescent channel before the pixel intensity levels were auto-contrast corrected and down-converted to an 8-bit RGB color image.  Using the cell perimeter coordinates generated with the Google Maps web-based UI, we extracted the intensity signal from the uncorrected F-actin channel.  We then set a threshold intensity level $I$ and calculated the F-actin density by adding all fluorescence signal above $I$ and dividing by the area of the pixels with intensity above $I$.  For each multinucleated cell, this produced a ``high-concentration'' measurement of the F-actin density.  Likewise, we performed an analogous measurement to calculate F-actin density in the ``low-concentration'' regions where the intensity level was below $I$.  We boot-strapped the data 11-fold by varying $I$ from 45\% to 55\% of the maximum intensity value in increments of 1\%.  In total, this produced 28,688 measurements of the F-actin density $\phi$ and averaged out the effect of specific intensity thresholds defining high and low concentration regions.  Of all cells and all measurements, 0.72\% of the data were discarded due to imaging artifacts from overlapping cells in high-density regions leaving 28,484 data points for constructing a phase diagram.  

Plotting the inverse size $1/A$ for each cell against its corresponding measurements of $\phi$ led to a dense mapping of points in phase space.  This dense map of points defines the occupied and unoccupied space that show a possible critical point and miscibility gap-like region (Fig.~4C,D).  Binning phase space enabled us to calculate the average phenotype by assigning a number to the cell classifications identified with unsupervised machine learning.  A uniform F-actin distribution was assigned 1, F-actin clusters were assigned 2, and the F-actin belt was assigned a 3.  For each bin, we calculated the mean (Fig.~4C) and the standard deviation (Fig.~4D) of these numerical values.

\section*{Experimental reproducibility}
All measurements of cell size $A$, number of nuclei $N$, and F-actin density $\phi$ are, in principle, possible to acquire using conventional manual imaging approaches.  The use of MAP imaging technology enables a more streamlined and automated processes that lowers the bar to accessibility and greatly reduces time requirements.  As such, we believe our main experimental results have a high degree of experimental reproducibility.

\section*{Statistical analysis}
Fits of model equations to empirical data were performed with MATLAB's curve fitting tool using the Trust-Region algorithm and default tolerances.  Quality of fits were reported with the adjusted coefficient of determination $R^2$.  In all cases, the difference between the ordinary and adjusted $R^2$ was less than 0.01.  Statistical tests for significance were calculated with an $N$-way analysis of variance (ANOVA) to ensure multiple simultaneous group comparisons were corrected for increased rates of false-positives.

\section*{Software availability}
An online demonstration of the Molecular Atlas Platform is available at: \url{www.molecular-atlas.net}.  This software, including all MATLAB analysis code, is freely available upon request for non-commercial uses.

\clearpage

\begin{figure*}
\includegraphics[scale=1]{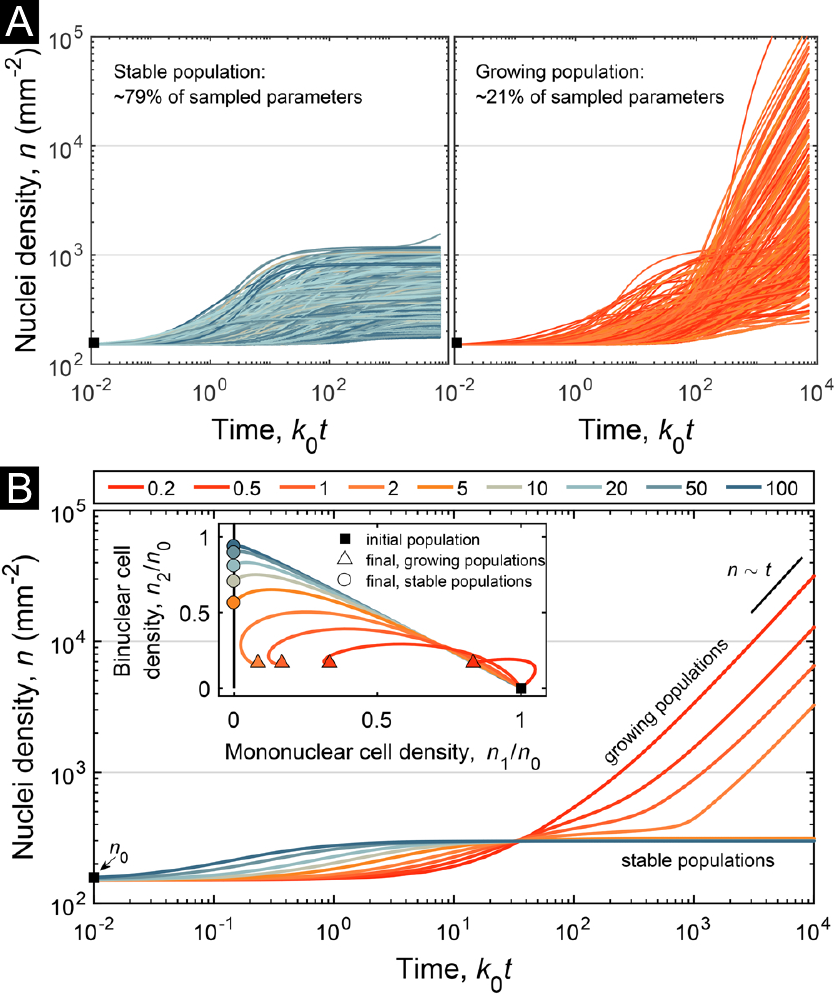}
\caption{Rate-equation modeling shows two stable states for the effects of fusion on population dynamics. (A) Stochastically sampling 30,000 sets of rate constants and simulating the population dynamics shows the total nuclei density either stabilizes due to a loss of precursor cells or grows linearly in time.  Here, we have normalized time by the mononuclear cell division rate $k_0$, which was also randomly generated in each run.  (B) A simplified version of the rate equation model enables precise testing of the balance between the mononuclear cell division rate $k_0$ and the mononuclear cell fusion rate $k_{11}$.  Numerical values in legend correspond to the ratio $k_{11}/k_0$ and show a clear transition between $k_{11}/k_0 = 2$ and 5 where the system goes from a growing population due to a finite number of mononuclear precursor cells $(n_1 > 0)$ to a stable population due to a complete loss of precursor cells $(n_1 = 0)$.  Inset shows phase portrait where the population densities $n_1$ and $n_2$ are normalized relative to the initial population density $n_0$.  Different trajectories from the same initial population either lead to a stable fixed point in phase space (triangles) or crash out on the $n_1 = 0$ line of fixed points (circles).  Notice that the location of the fixed point varies with the ratio $k_{11}/k_0$.}
\label{figS1}
\end{figure*}


\begin{figure*}
\includegraphics[scale=1]{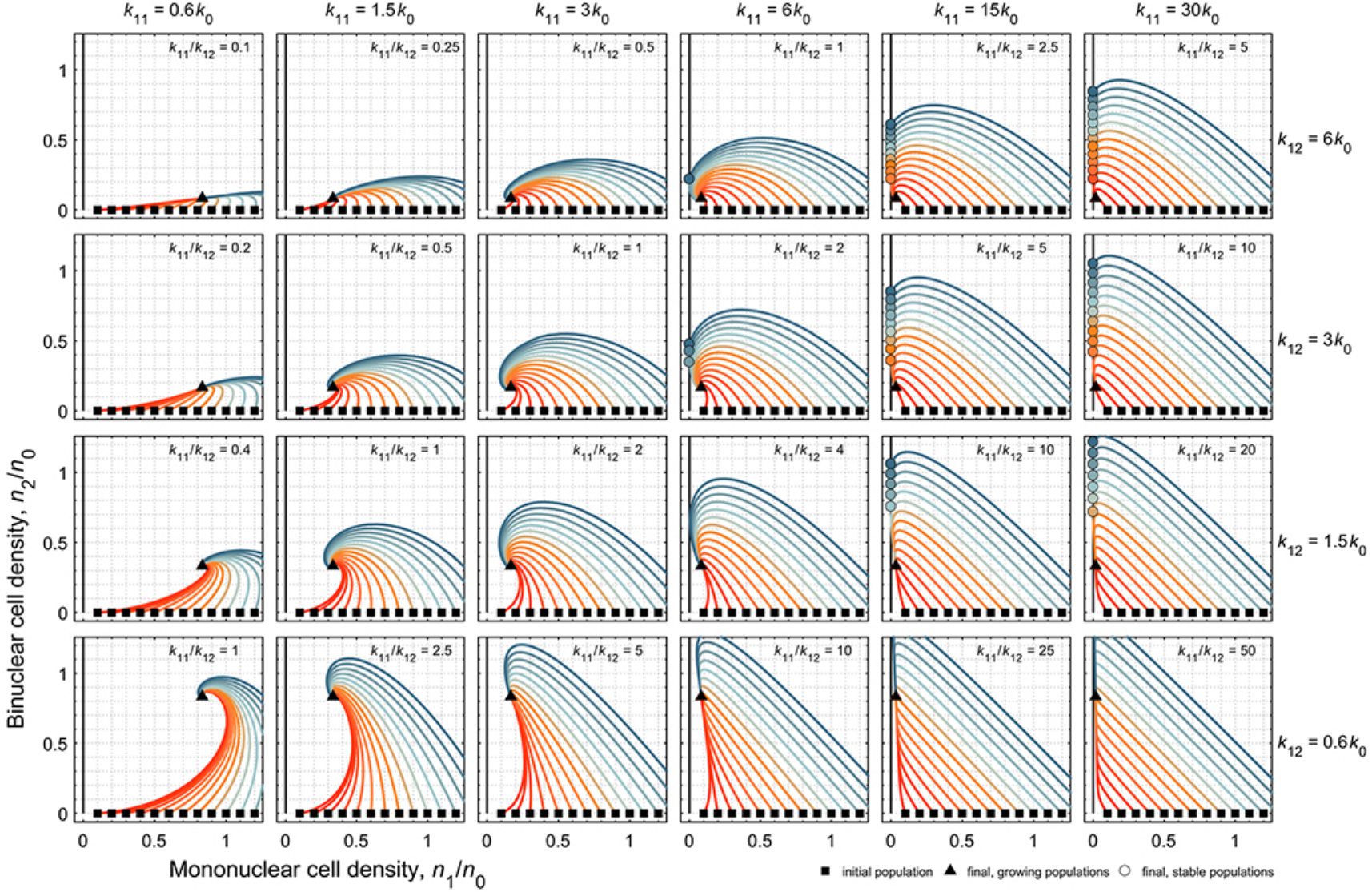}
\caption{Extended phase portrait analysis of simplified fusion population dynamics model. Varying the fusion rates $k_{11}$ and $k_{12}$ relative to the division rate changes where the fixed-point position (triangle) is in phase space.  Note that this fixed-point corresponds to a stable precursor population with $n_1$ remaining constant.  Varying the initial population $n_0$ (squares) changes the specific trajectory path.  Some trajectories stabalize, while others experience a population crash when $n_1$ goes to 0 (circles).  The line $n_1 = 0$ corresponds to a fixed line in phase space where the total population of cells remains constant because there are no precursor cells left to divide and multiply.  }
\label{figS2}
\end{figure*}


\begin{figure*}
\includegraphics[scale=1]{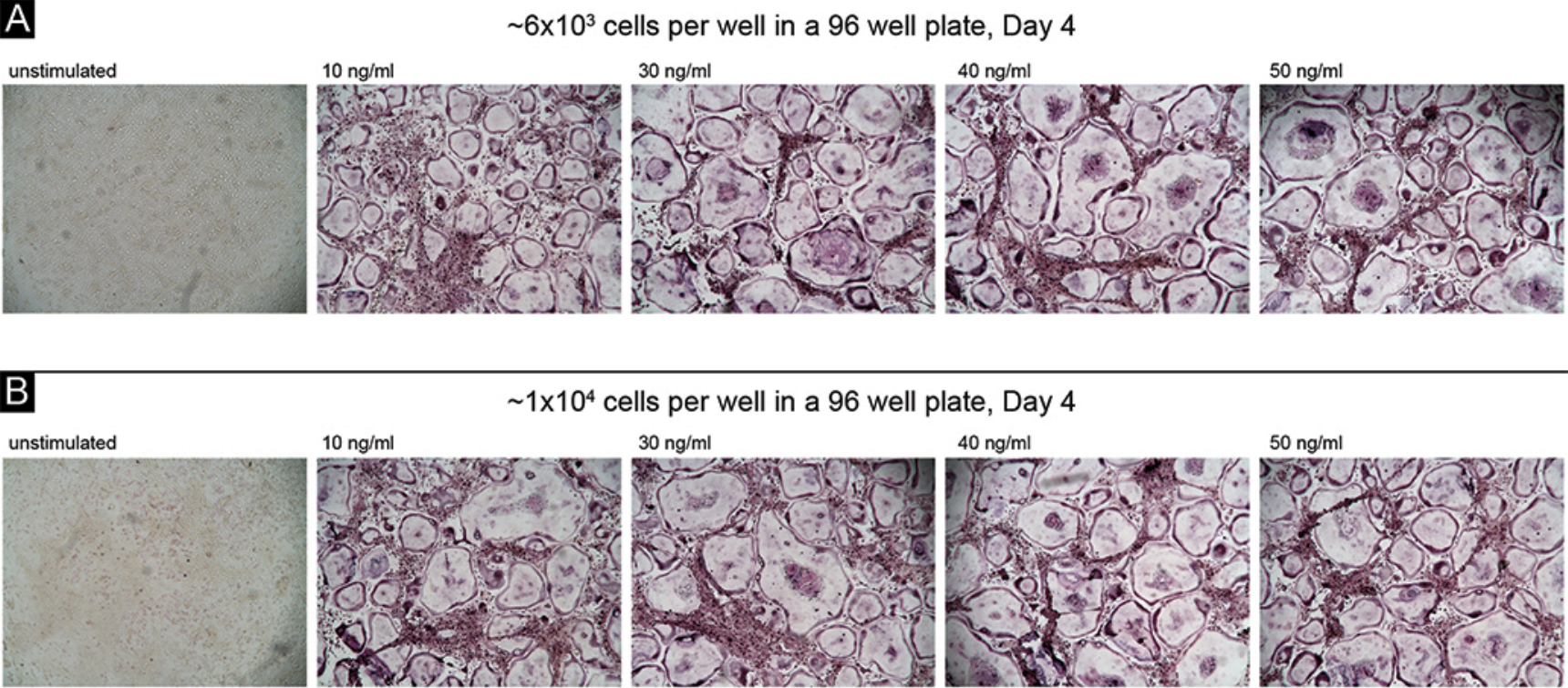}
\caption{RANKL stimulation triggers fusion of precursor macrophage cells. (A) Optical microscopy images with TRAP stained cells plated at $6 \times 10^3$ cells per well.  The unstimulated macrophages (left) show no TRAP activity.  Introducing RANKL functions as an on/off switch that causes these precursor cells to fuse into osteoclasts and giant multinucleated cells.  Increasing the RANKL concentration from 10 ng/ml to 50 ng/ml indicates that fusion activity is broadly self-consistent despite this 5-fold increase in signaling ligand concentration.  (B) Plating precursor macrophage cells at $10^4$ cells per well shows a similar on/off behavior with RANKL stimulation and broad consistency across RANKL concentrations. }
\label{figS3}
\end{figure*}


\begin{figure*}
\includegraphics[scale=1]{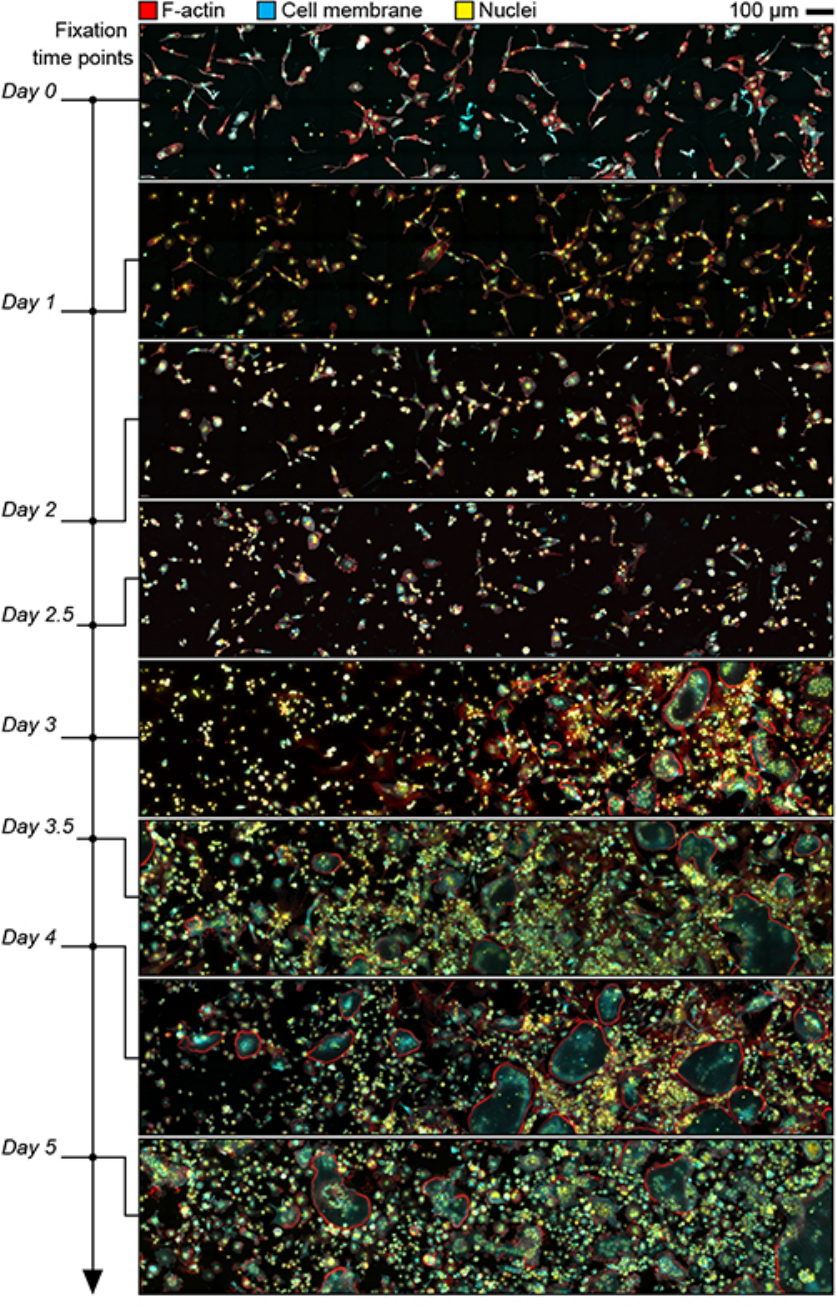}
\caption{Empirical and theoretical population dynamics. These eight images are partial fields of view from the larger gigapixel-scale images acquired experimentally.  The total width, height, and number of $z$-planes varied somewhat from sample-to-sample, causing the acquisition time to vary between 8 and 24 hours.  }
\label{figS4}
\end{figure*}

\begin{figure*}
\includegraphics[scale=1]{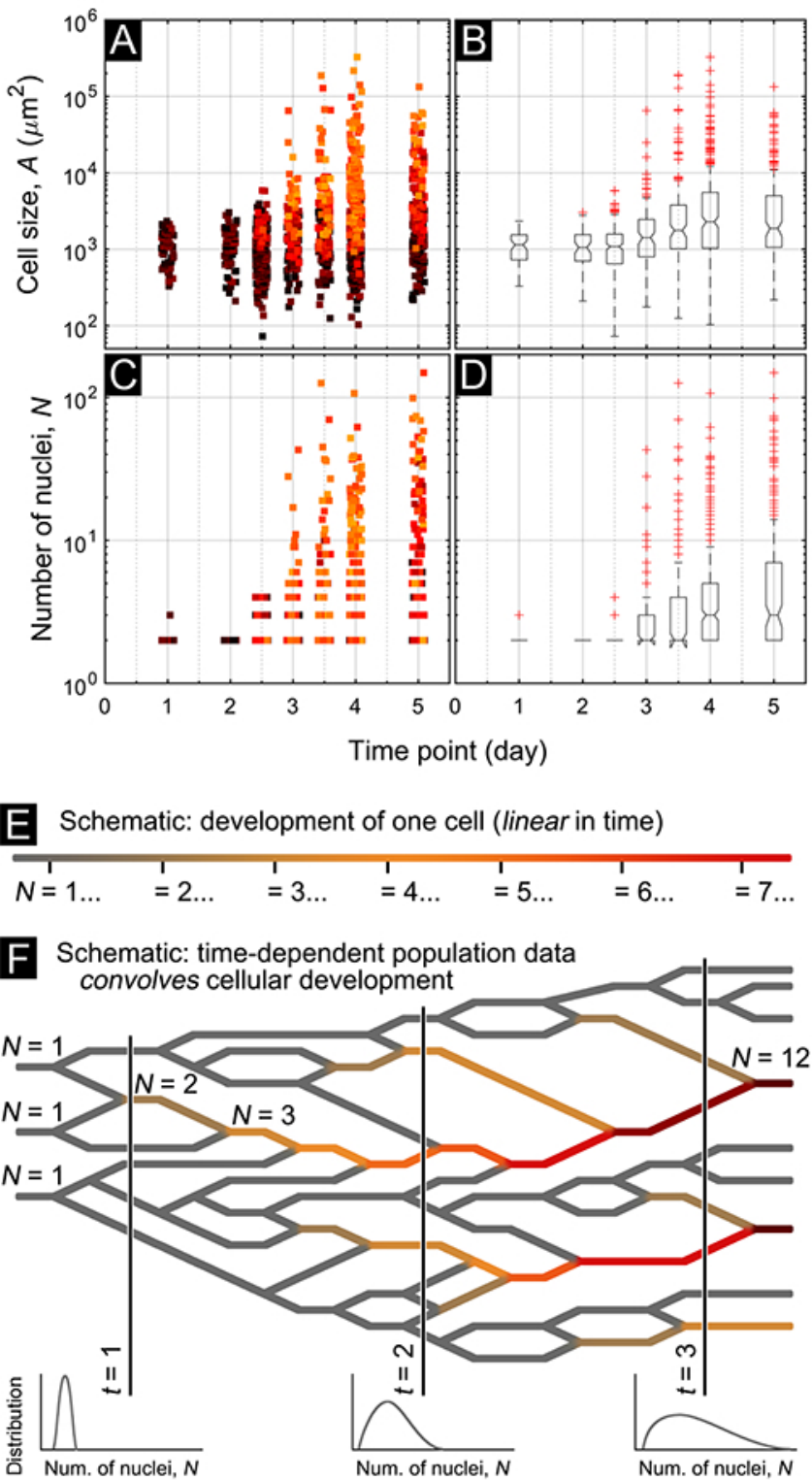}
\caption{Parametrization by time convolves measurements and obscures trends.  (A) Multinucleated cell size measurements plotted at various fixation time points show a general trend to increase, both in mean and fluctuation.  (B) Box and whisker plots excluding outliers indicate the difficulty in showing statistical significance.  When the notches of two columns are non-overlapping, then the medians are different with $p < 0.05$.  When notches overlap, then the two columns show no statistically significant differences. (C) Similarly, the distribution for the number of nuclei broadens over time. (D) Box and whisker plots excluding outliers again fail to show strong statistically significant results due to the broadened distributions.  As with panel (B) non-overlapping notches indicate data with statistically significant differences ($p < 0.05$).  (E) We can schematically indicate development associated with increasing number of nuclei by a linear color gradient.  (F) A lineage tree (time increases left-to-right, number of nuclei indicated by color as in (E)) with three initial mononuclear precursors (gray), that divide into two daughter cells (forks) lead to a series of fusion events (mergers).  While the number of nuclei increases linearly for individual multinucleated cells (reading diagram horizontally, left-to-right), snap-shots at specific times points (reading diagram vertically) show how quantitative metrics inevitably measure distributions that broaden in time (compare time points $t = 1, 2,$ and $3$).  }
\label{figS5}
\end{figure*}


\begin{figure*}
\includegraphics[scale=1]{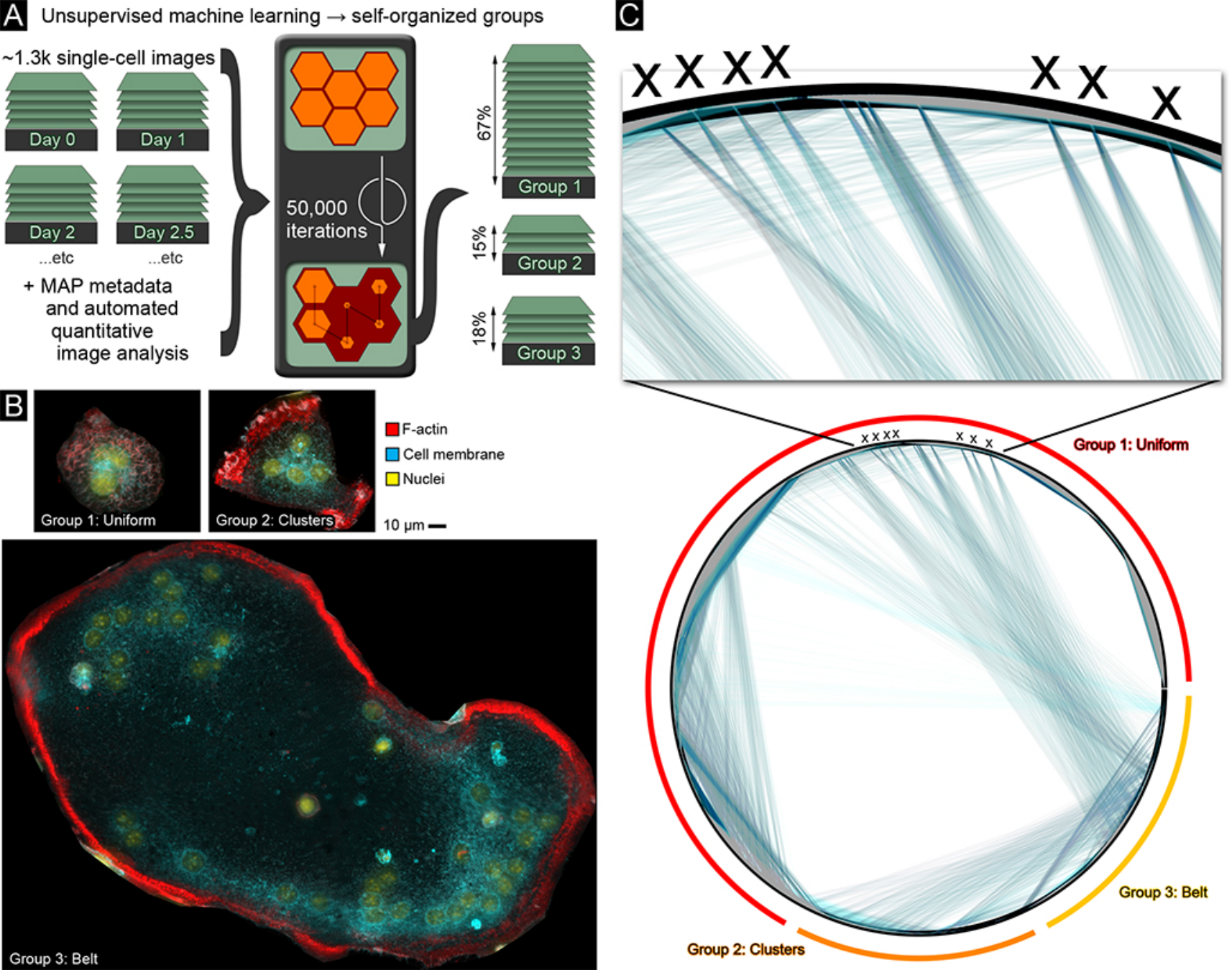}
\caption{Schematic of unsupervised machine learning. (A) Diagram of computational work-flow starting with quantitative image analysis applied to experimental image data sets. This panel illustrates a $2 \times 3$ neural network that undergoes iterative training to identify natural groupings of the data.  Percentages indicate the portion of input data classified into their respective groups.  (B) Example images from each of the three groups and the corresponding descriptions known from osteoclastogenesis.  Notice the F-actin has a distinct reorganization associated with the increased cell size.  Cells adjacent to these three examples have been masked using the annotation data for the cell perimeter.  (C) Schemaball representation of the adjacency matrix from 50 trained self-organized map neural networks averaged together.  Groups are indicated in the outer-most red, orange, and yellow arcs, individual cells are arranged as densely packed nodes forming the black outer ring, and colored lines within the schemaball connect cells that appear together in 85\% (blue) or more (decreasing saturation indicates higher match rates) of the trained neural networks.  The vast majority of connecting lines are along the perimeter and form the nearly solid color bands.  Approximately 4\% of the connections are from cells in different groups and are therefore misclassified by the trained neural network in (A).  Seven example cells in Group 1 are labeled (black $x$'s), and the zoomed inset shows how these cells frequently cluster together with cells in Group 3.}
\label{figS6}
\end{figure*}


\begin{figure*}
\includegraphics[scale=1]{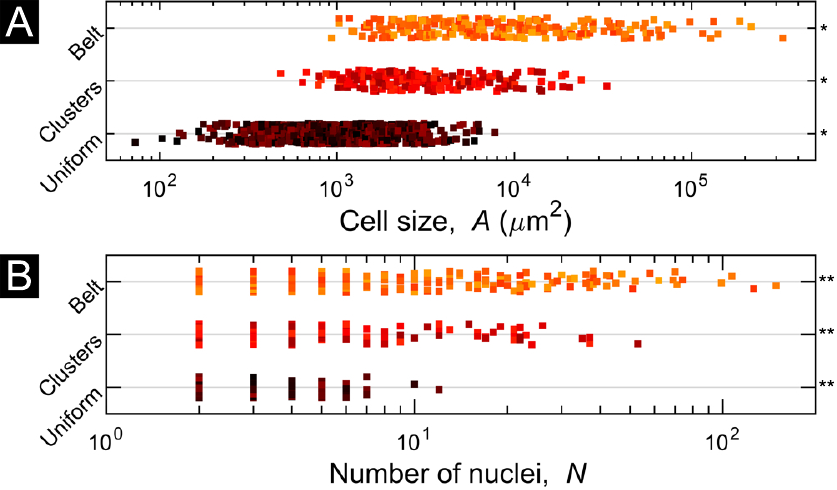}
\caption{Unsupervised machine learning identifies statistically significant groups. (A) The three groups identified by trained neural networks have differences in multinucleated cell size $A$, and (B), number of nuclei $N$.  Here, * indicates $p < 10^{-47}$ and ** indicates $p < 10^{-66}$.  }
\label{figS7}
\end{figure*}

\end{widetext}

\clearpage

\end{document}